\documentclass[format=acmsmall, review=false, screen=true]{acmart}

\usepackage{graphicx,amsmath,mathrsfs,epsf}
\usepackage{multirow}
\usepackage{tabularx}
\usepackage{stfloats}
\usepackage{setspace}
\usepackage{bbold}
\usepackage{bm}
\usepackage{subcaption}
\usepackage{bigstrut,array,multirow,tabularx}
\usepackage{tikz}
\usepackage{xcolor}
\usepackage{pgfplots}
\usepackage{dsfont}
\usepackage{bbm}
\usepackage{subcaption}
\usepackage[ruled, vlined, linesnumbered]{algorithm2e}

\newtheorem{definition}{Definition}

\def \log{\operatorname{log}}

\def \exp{\operatorname{exp}}

\def \be {\begin{eqnarray}}
\def \ee {\end{eqnarray}}
\def \ben {\begin{eqnarray*}}
\def \een {\end{eqnarray*}}

\newcommand{\argmin}{\mathop{\mathrm{arg\,min}}}
\newcommand{\argmax}{\mathop{\mathrm{arg\,max}}}

\newcommand{\N}{\mathbbmss{N}}
\newcommand{\R}{\mathbbmss{R}}

\acmJournal{TKDD}
\acmVolume{0}
\acmNumber{0}
\acmArticle{0}
\acmYear{0000}
\acmMonth{0}

\setcopyright{acmlicensed}

\acmDOI{0000001.0000001}

\begin{document}

\title[Community Detection in Partially Observable Social Networks]{Community Detection in Partially Observable Social Networks}
\author{Cong Tran}
\affiliation{%
  \institution{Department of Computer Science and Engineering, Dankook University}
  \city{Yongin}
  \country{Republic of Korea}}
\affiliation{%
  \institution{Machine Intelligence \& Data Science Laboratory, Yonsei University}
  \city{Seoul}
  \country{Republic of Korea}}
\email{trancong208@gmail.com}

\author{Won-Yong Shin}
\affiliation{%
  \institution{School of Mathematics and Computing (Computational Science and Engineering), Yonsei University}
  \city{Seoul}
  \country{Republic of Korea}}
\email{wy.shin@yonsei.ac.kr}

\author{Andreas Spitz}
\affiliation{%
  \institution{Department of Computer and Information Science, University of Konstanz}
  \city{Konstanz}
  \country{Germany}}
\email{andreas.spitz@uni-konstanz.de}
\begin{abstract}
The discovery of community structures in social networks has gained significant attention since it is a fundamental problem in understanding the networks' topology and functions. However, most social network data are collected from partially observable networks with {\em both missing nodes and edges}. In this paper, we address a new problem of detecting {\em overlapping} community structures in the context of such an {\em incomplete} network, where communities in the network are allowed to overlap since nodes belong to multiple communities at once. To solve this problem, we introduce \textsf{KroMFac}, a new framework that conducts community detection via regularized {\em nonnegative matrix factorization (NMF)} based on the {\em Kronecker graph} model. Specifically, from an inferred Kronecker generative parameter matrix, we first estimate the missing part of the network. As our major contribution to the proposed framework, to improve community detection accuracy, we then characterize and select {\em influential} nodes (which tend to have high degrees) by ranking, and add them to the existing graph. Finally, we uncover the community structures by solving the regularized NMF-aided optimization problem in terms of maximizing the likelihood of the underlying graph. Furthermore, adopting normalized mutual information (NMI), we empirically show superiority of our \textsf{KroMFac} approach over two baseline schemes by using both synthetic and real-world networks.
\end{abstract}

%
%



\maketitle

\section{Introduction}\label{sec:1}
\subsection{Backgrounds}
Real-world networks extracted from various biological, social, technological, and information systems usually contain inhomogeneities that reveal a high level of hierarchical and structural properties. Research on community detection, which is one of the most important tasks in network analysis, has thus become crucial in understanding the fundamental features (e.g., topology and functions) of these networks~\cite{fortunato2010community}. In general terms, communities can be regarded as the sets of points that are ``close" to each other with respect to a predefined measure of distance or similarity. Since applications of community detection are diverse, there exist a variety of graph-theoretic approaches~\cite{leskovec2010empirical} that conduct optimization based on measures such as modularity~\cite{newman2006modularity} and conductance~\cite{leung2009towards}, whose performance depends heavily on network topology. 

On the one hand, community detection algorithms for online social networks should be designed by taking into account their inherently overlapping and imprecise nature, since community memberships in social networks are allowed to overlap as nodes belong to multiple clusters at once~\cite{xie2013overlapping}. The extraction of such overlapping communities is known to be more challenging than non-overlapping community detection due to higher complexity and higher computational demands.

In practice, on the other hand, most social network data are collected from partially observable networks with {\em both missing nodes and edges}~\cite{kossinets2006effects}, which further complicates the detection of communities. For example, due to limited resources, a person or an organization may be allowed to obtain only a subset of data within a specific geographic query region. This is further compounded due to privacy settings specified by the users that may partially or entirely hide some of their traces or friendships~\cite{acquisti2015privacy}. For example, 52.6\% of Facebook users in New York City hid their friend lists in June 2011~\cite{dey2012facebook}. Such types of incomplete networks constitute a severe obstacle for topology-based optimization methods in detecting the true community structures. Surprisingly, while some research exists into the recovery of edges and nodes in such incomplete networks, the problem of community detection under such conditions and its solutions have not yet been investigated.
\subsection{Motivation and Main Contributions}
In this work, we formulate a new problem of detecting overlapping community structures in the context of such an {\em incomplete} network in which some of the nodes and edges are missing. To solve the problem, we present \textsf{KroMFac}, a new framework that intelligently combines network completion and community detection methods into one {\em unified framework}, which is the first attempt in the literature. To this end, \textsf{KroMFac} first estimates the missing part of the network using a Kronecker generative parameter matrix acquired under the {\em Kronecker graph} model~\cite{kronem}, which basically differs from the well-known link prediction task in machine learning since node labels would never be acquired by link prediction. Our important contribution to the proposed framework is based on the insight that including the entirety of recovered nodes and edges in the existing graph may be detrimental to enhancement of community detection accuracy. This is because adding more recovered nodes and edges would cause the inference errors to accumulate. To address this problem, we characterize and select {\em influential nodes} by {\em centrality ranking}, which tend to have high degrees, in the effort of limiting the accumulated errors in our model. Finally, we perform community detection using a state-of-the-art algorithm along with the recovered graph in which influential nodes are added. In this study, rather than adopting recently developed deep learning-based approaches,  we focus on {\em nonnegative matrix factorization (NMF)}-based community detection due to its considerably better scalability while guaranteeing satisfactory performance. More precisely, we formulate a regularized NMF-aided optimization problem in terms of maximizing the likelihood of the underlying graph to discover the community structures. After solving the problem, we assign nodes to communities depending on the values of each entry in the factorized affiliation matrix. 

Adopting normalized mutual information (NMI) as a popular information-theoretic performance metric, we empirically verify the superior performance of our proposed approach over two baselines that 1) do not infer missing nodes and edges (Baseline 1) and 2) leverage completion of the entire network (Baseline 2). In summary, our main contributions are five-fold and summarized as follows:
\begin{itemize}
\item design of a new framework, named \textsf{KroMFac}, that intelligently combines network completion and community detection in our incomplete network; 
\item formulation of a regularized NMF-aided joint optimization problem; 
\item characterization and selection of influential nodes via ranking, which play a vital role in improving community detection accuracy; 
\item validation of our \textsf{KroMFac} approach through intensive experiments based on parameter search using both synthetic and real-world datasets;
\item analysis and empirical validation of the computational complexity.
\end{itemize}

Our framework takes an important first step towards establishing a new line of research and towards a better understanding of jointly conducting both network recovery and community detection in partially observable networks.
\subsection{Organization}
The remainder of this paper is organized as follows. In Section~\ref{sec:2}, we summarize significant studies that are related to our work. In Section~\ref{sec:3}, we explain the methodology of our work, including the problem definition and the overview of our \textsf{KroMFac} framework. Section~\ref{sec:4} describes implementation details of the proposed \textsf{KroMFac} framework. Experimental results are provided in Section~\ref{sec:5} with comparison to two baseline approaches. Finally, we summarize the paper with some concluding remarks in Section~\ref{sec:6}.

\subsection{Notations}
Throughout this paper, $\mathbb{R}$ and $\mathbb{P}(\cdot)$ indicate the field of real numbers and the probability, respectively. Unless otherwise stated, all logarithms are assumed to be to the base $e$. Table~\ref{table:notation} summarizes the notations used in this paper. These notations will be formally defined in the following sections when we introduce our network model and technical details. 
\begin{table}[]
\centering
\caption{Summary of notations.}
\label{table:notation}
\begin{tabular}{l|l}
\hline
Notation               & Description                                 \\ \hline
$G$                    & observable graph                            \\
$V$                    & set of observable nodes                     \\
$E$                    & set of observable edges                     \\
$V_M$                  & set of missing nodes                        \\
$E_M$                  & set of missing edges                        \\
$N$                    & number of observable nodes                  \\
$M$                    & number of missing nodes                     \\
$G'$                   & true graph                                  \\
${\bf F}$              & affiliation matrix                          \\
$C$                    & number of communities                       \\
${\bf A}$              & adjacency matrix of the observable graph $G$   \\
${\bf A'}$             & adjacency matrix of the true graph $G'$         \\
$i$                    & number of recovered nodes                   \\
$R^{(i)}$              & recovered graph after connecting $i$ nodes    \\
${\bf A}_R^{(i)}$      & adjacency matrix of the recovered graph $R^{(i)}$       \\
${\bf Z_1}$            & matrix containing links between recovered nodes and existing nodes            \\
${\bf Z_2}$            & matrix containing links between between recovered nodes            \\
$H$                    & number of influential nodes                 \\
$\lambda$  & regularization parameter				\\
$\bm{\Theta}$          & Kronecker parameter matrix                  \\
$\bm{\Theta}_{\text{init}}$          & initialized Kronecker parameter matrix                  \\
$K$                      & number of Kronecker products                   \\
$\text{Cen}$         & degree centrality                    \\
$\mathcal{D}$          & loss function                               \\
$\epsilon$             & threshold for determining influential nodes        \\
$\psi$                 & set of communities                          \\
$\delta$               & threshold determining communities              \\
${\bm r}$              & ranking vector                              \\
\hline
\end{tabular}
\end{table}
\section{Related Work}\label{sec:2}
The framework that we propose in this paper is related to four broader areas of research, namely community detection in graphs, detection of overlapping communities, community detection in incomplete networks with missing edges, and network completion in social networks.

{\bf Community detection in graphs.} Since research into community detection in complex networks constitutes a very active field, there are many efforts devoted to community detection in graphs. The most popular techniques include modularity optimization~\cite{newman2006modularity}, stochastic block models~\cite{sbm}, spectral graph-partitioning~\cite{newman2013spectral}, clique percolation~\cite{du2007community}, clustering~\cite{chen2012dense}, and label propagation~\cite{raghavan2007near,tkdd1}. However, these techniques focus on graphs in which nodes can be partitioned into disjoint communities and do not address the inherent overlapping nature of community structures in many real-world networks. 

{\bf Detection of overlapping communities.} To cope with this contrast, a recently emerging topic covers the detection of overlapping communities by investigating the structural properties of such communities, especially in the case of social networks~\cite{agm,papadopoulos2012community}. As the computational complexity increases drastically for the recovery of overlapping communities instead of partitioned communities, the research has focused either on detecting communities based on local expansion~\cite{whang2016overlapping,tkdd2} or on employing scalable techniques such as NMF~\cite{bigclam,BNMF,newnmf} and label propagation~\cite{COPRA,SLPA}. More recently, sophisticated network embedding-based approaches for overlapping community detection were presented \cite{linkblack,linkembedding,vgraph,gnn1,gnn2,TKDDgnn}. Specifically, overlapping communities were found via clustering in a link embedding space \cite{linkblack,linkembedding}. To jointly perform community detection and node representation learning, vGraph \cite{vgraph} was developed based on  a generative model where the community membership acts as a latent variable generating neighbors for each node. In \cite{gnn1}, communities were detected by employing graph neural networks with a non-backtracking edge adjacency structure to capture complex network patterns. In \cite{gnn2}, a Bernoulli--Poisson probabilistic model was combined with a graph convolutional network to perform community detection. In \cite{TKDDgnn},  clustering (i.e., detecting communities) and node representation were jointly optimized in attributed networks.

{\bf Community detection in incomplete networks.} Recently, research on community detection in incomplete networks with {\em missing edges} has attracted wide attention due to a lack of information caused by users' privacy settings and limited resources. Most of these studies predict the missing links between nodes based on the incorporated additional information~\cite{yang2013community} or the similarity of topological structures~\cite{yan2011finding,yan2012detecting}, and then discover communities in the underlying recovered networks. In~\cite{zhou2015infinite}, a hierarchical gamma process infinite edge partition model was presented to detect communities and recover missing edges in parallel. Furthermore, several studies address the case when the underlying network structure is not available; in such a case, communities can be extracted directly from several sources of raw information, such as influence cascades  \cite{comcascades}  and time-series  data \cite{comtimedata}.

{\bf Network completion in social networks.} In addition to the studies on community detection, network completion thus plays an important role in our research since it should precede the community detection process. As the most influential study, KronEM, an approach based on Kronecker graphs to solving the problem  of discovering both missing nodes and edges was suggested by Kim and Leskovec~\cite{kronem}, where the expectation maximization (EM) algorithm is applied. For cases in which only a small number of edges are missing, vertex similarity~\cite{chen2011capturing} was shown to be useful in recovering the original networks. As another topic of network completion, the missing node identification problem was introduced in~\cite{misc}, which aims to {\em cluster} the available connections between a number of unidentified placeholders and observable nodes. Aside from KronEM \cite{kronem}, no other studies have tackled the problem where a node and its associated edges are entirely missing, which is the setting that we consider in this study.

Despite these contributions, there has been no prior work in the literature that combines the contexts of community detection in incomplete social networks with the recovery of both missing nodes and edges. In the following, we therefore present such an approach that seamlessly integrates the recovery of missing parts of a network with subsequent community detection on the recovered network, while benefiting from a  resulting more complete community structure.

\section{Methodology}\label{sec:3}

As basis for the algorithms in Section~\ref{sec:4}, we discuss network fundamentals, formalize the problem definition, and then introduce the generative graph model for network completion, our node selection strategy, and the community detection method.

\subsection{Problem Definition}\label{sec:3a}
\subsubsection{Network Model and Basic Assumptions}
Let us denote the partially observable network as $G = (V, E)$, where $V$ and $E$ are the set of vertices and the set of edges, respectively. The network $G$ with $N=|V|$ nodes can be interpreted as a subgraph taken from an underlying true social network $G' = (V~\cup~V_M,~E~\cup~E_M)$, where $V_M$ is the set of unobservable nodes and $E_M$ is the set of unobservable edges. If we assume $G'$ to be a scale-free network, then the degree distribution of $G'$ can be approximated as $\mathbb{P}(k) \sim k^{-\gamma}$, where the probability $\mathbb{P}(k)$ of a node in the network is inversely proportional to its degree $k$ raised to the power of an exponent parameter $\gamma$.\footnote{The degree distribution can be estimated via least squares approximation just by taking at most 1\% of the samples using a sublinear approach as indicated in~\cite{estimateexponent}.} 
While not all real-world social networks necessarily follow a power-law distribution~\cite{clauset2009powerLaw}, fitting a power-law model to the long tail of the distribution is usually sufficient for practical applications. Therefore, other types of networks following a heavy-tailed degree distribution can also serve as suitable input for our work. In real-world social networks, the nodes and edges of the network $G'$ correspond to users and their relationships, respectively, with little additional information being available. In the following, we thus consider both $G$ and $G'$ to be {\em undirected} unweighted networks.
\begin{figure}[!t]
    \begin{center}
        \begin{subfigure}[]{0.25\textwidth}
		\centering
            \includegraphics[height=2cm]{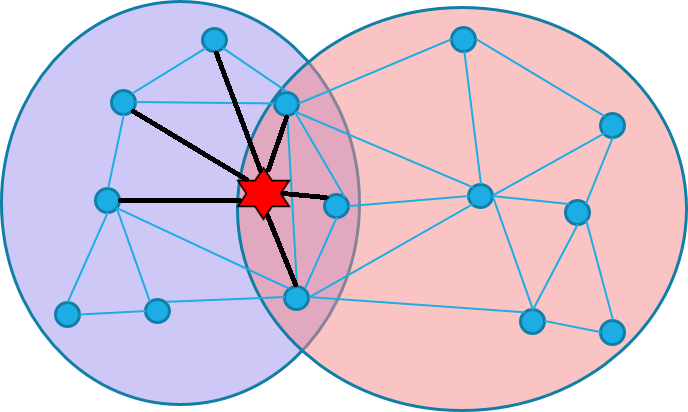}
            \caption{{Before node deletion}}
            \label{fig:before}
        \end{subfigure}%
        \begin{subfigure}[]{0.25\textwidth}
		\centering
            \includegraphics[height=2cm]{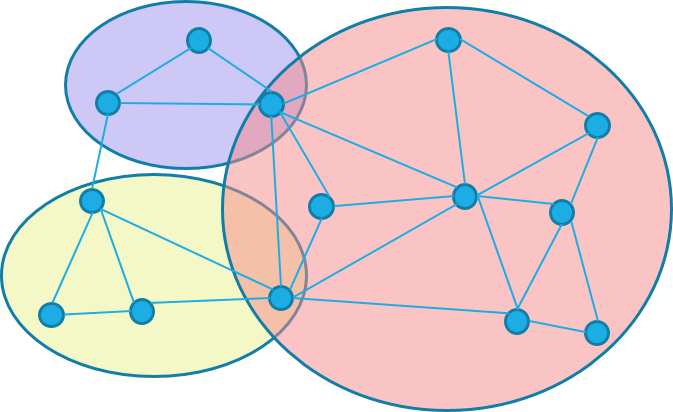}
            \caption{{After node deletion} }
            \label{fig:after}
        \end{subfigure}
    \end{center}
    \caption{An example that illustrates the difference between the community structures before and after an influential node (star) and its incident edges (black) have been deleted from the graph. Potential communities are depicted with different colors.}
    \label{fig:changestructure}
\end{figure}
Furthermore, we assume that the number of missing nodes $M = |V_M|$ is either known or can be approximated by standard methods for estimating the size of hidden or missing populations~\cite{estimatepopulation}. To detect overlapping communities, we assume that social networks follow the {\em affiliation graph model (AGM)}~\cite{agm}, which states that the more communities a pair of nodes shares, the higher the probability that these two nodes are connected. The number of communities in the network is denoted by $C$. The AGM can be represented by a non-negative weight affiliation matrix ${\bf F} \in \R^{(N + M) \times C}$ such that each element ${\bf F}_{uc}$ represents the degree of membership of a node $u \in (V\cup V_M)$ to the community $c$. The probability $\mathbb{P}(u,v)$ of a connection between two nodes $u$ and $v$ then depends on the value of ${\bf F}$ and is given by $\mathbb{P}(u,v) = 1 - $ exp$(-{\bf F}_{u} {\bf F}^{\top}_{v})$, where ${\bf F}_{u} \in \R^C$ and ${\bf F}_{v} \in \R^C$ are the row vectors that correspond to nodes $u$ and $v$, respectively~\cite{bigclam}. 
\begin{figure*}[t]
    \begin{center}
            \includegraphics[height=3.5cm]{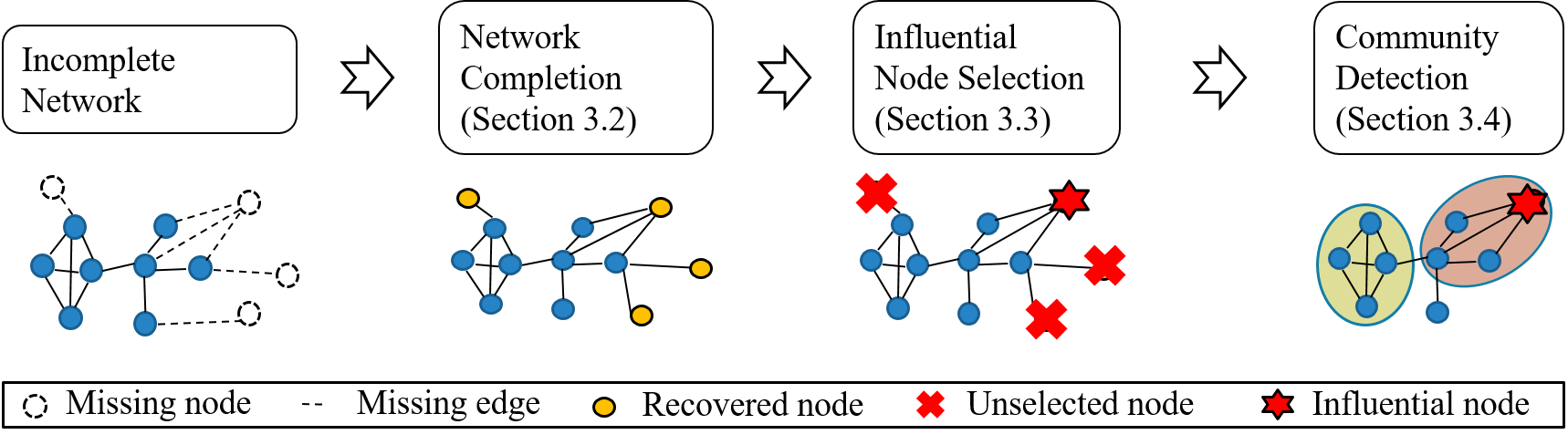}
	\hspace{2mm}
            \caption{{The schematic overview of our \textbf{\textsf{KroMFac}} framework.} }
            \label{fig:methodology}
    \end{center}
\vspace{-0.15in}
\end{figure*}
\subsubsection{Problem Formulation} As illustrated in Figure~\ref{fig:changestructure}, the network structures of partially observable networks are potentially distorted significantly due to the effect of both missing nodes and edges. As a result, methods established for detecting communities may appear to perform well on the partially observable networks but are not effective in extracting the true community structures of the underlying true network. 
The recovery of overlapping communities in such incomplete, partially observable networks has not been investigated before in the literature. To address this task, we thus present \textsf{KroMFac}, a novel framework for recovering a partially observable network and then discovering the overlapping community structures of the recovered underlying graph. To this end, we first recover missing nodes and edges, which is equivalent to filling in the missing part of the binary adjacency matrix ${\bf A}' \in \{0,1\}^{(N+M) \times (N+M)}$ of the graph $G'$ based on the topological information of the observable matrix ${\bf A}$ (refer to Section~\ref{sec:3b}). This inference of missing parts of the network is not without risk since adding more recovered nodes and edges may also accumulate more errors. However, many such nodes and edges may not be very relevant to the subsequent community detection. This motivates us to propose a node selection strategy that aims to characterize and include only a small number of nodes that have a high impact on the community detection. Specifically, we need to {\em selectively} recover nodes. We start by formally defining the adjacency matrix that is acquired after the iterative addition of nodes (and their edges) according to an importance ranking strategy.

\begin{definition}\label{def:1}
Let $R^{(i)}$ be the selectively recovered graph formed by connecting $i \in\{0,1,\cdots,M\}$ nodes to the existing graph $G$ according to a predefined selection order. Based on the fact that $G$ and $R^{(i)}$ correspond to ${\bf A} \in \{0,1\}^{N \times N}$ and ${\bf A}_R^{(i)}  \in \{0,1\}^{(N + i )\times (N + i)}$, respectively, ${\bf A}_R^{(i)}$ can be written as the following partitioned block matrix:
\[
{\bf A}_R^{(i)}=
  \begin{bmatrix}
    {\bf A} & {\bf Z}_1 \\
    {\bf Z}_1^\top & {\bf Z}_2
  \end{bmatrix},
\]
where the matrix ${\bf Z}_1 \in \{0,1\}^{N \times i}$ contains the links between recovered nodes and existing nodes and the matrix ${\bf Z}_2 \in \{0,1\}^{i \times i}$ contains the links between recovered nodes. 
\end{definition}

By definition, if we select the top $i$ nodes, then we obtain a unique matrix ${\bf A}_R^{(i)}$. As special cases, it follows that ${\bf A}_R^{(0)} = {\bf A}$ and ${\bf A}_R^{(M)} = {\bf A}'$. To limit the accumulated errors to a certain level in our model, we only take into account top $H\in\{0,1,\cdots,M\}$ nodes in the ranked list, termed {\em influential nodes} (refer to Definition 2 in Section~\ref{sec:3c}).

The next step is the detection of communities, which is equivalent to estimating the affiliation matrix ${\bf F}\in \mathbb{R}^{(N+i)\times C}$. For each $i$ (i.e., the number of newly added nodes), estimation of the affiliation matrix ${\bf F}$ leads to a probabilistic approximation of the matrix ${\bf A}_R^{(i)}$ (refer to Section~\ref{sec:3d}). When $\hat{{\bf F}}$ has the highest chance to generate the graph $R^{(\hat{i})}$, we formulate a joint optimization problem as follows:
\begin{equation}\label{eq:1}
(\hat{{\bf F}}, \hat{i})=\argmax_{{\bf F} \geq 0, i \in\{0,1,\cdots,H\} } \log\mathbb{P}({\bf A}_R^{(i)}|{\bf F} ) + \lambda\log(i+1),
\end{equation}
which corresponds to a maximum log-likelihood problem with {\em regularization} for a given value of $i$.\footnote{Here, the superscript $\hat{}$ is used to indicate the optimal argument.} Here, $\log(i+1)$ indicates a regularization term, which is used to compensate the log-likelihood $\mathbb{P}({\bf A}_R^{(i)}|{\bf F} )$ reduced by increasing the size of the matrix ${\bf A}_R^{(i)}$ since the probability $\mathbb{P}({\bf A}_R^{(i)}|{\bf F})$ tends to decrease with the number of elements of ${\bf A}_R^{(i)}$; and the parameter $\lambda>0$ determines the impact of the regularization term and needs to be properly set according to the size of observable graphs (which will be specified in Section~\ref{sec:5c}). The role of $i$ in~(\ref{eq:1}) is especially important as it is the key factor in handling the total error of the recovered graph. 
The overall procedure of our approach is visualized in Figure~\ref{fig:methodology}.

\subsection{Generative Graph Model}\label{sec:3b}
For recovery of the true network structures, the missing part of the network can be inferred by investigating the connectivity patterns in the observable part. To this end, {\em generative} models for graphs have been developed. The two major generative graph models with this aim include the stochastic block model~\cite{sbm} and the Kronecker graph model~\cite{kronmodel}. For our research, we adopt the Kronecker graph model since it is scalable and can be used to efficiently model a probability distribution over the missing part of social networks~\cite{kronmodel}. Thus, we briefly describe the Kronecker graph model before proceeding to network completion.

The model is based on the Kronecker product of two graphs~\cite{kronproduct}. For two given adjacency matrices ${\bf A} \in \R^{m \times n}$ and ${\bf B} \in \R^{m' \times n'}$, the Kronecker product ${\bf A} \otimes {\bf B} \in \R^{mm' \times nn'}$ is defined as
\[
{\bf A} \otimes {\bf B}=
  \begin{bmatrix}
    a_{11}{\bf B}  & \dots & a_{1n}{\bf B}\\
    \vdots & \ddots & \vdots\\
    a_{m1}{\bf B} & \dots & a_{mn}{\bf B}
  \end{bmatrix},
\]
where $a_{uv}$ denotes the $(u,v)$th element of the matrix ${\bf A}$ for $u\in\{1,\cdots,m\}$ and $v\in\{1,\cdots,n\}$.
The Kronecker graph model is then defined by a Kronecker generative parameter matrix $\bm{\Theta} \in [0,1]^{N_0 \times N_0}$, where $N_0 \in \N$.\footnote{The parameter $N_0$ is typically set to two to model the structure of social networks~\cite{kronmodel}, but it can also be set to any integer so that there is no limit in the network size.} By Kronecker-powering the parameter $\bm{\Theta}$, we obtain increasingly larger and larger stochastic graph adjacency matrices. Since every entry of the matrix $\bm{\Theta}$ can be interpreted as a probability, the Kronecker graph model is then equivalent to a probability distribution of edges over networks.

For network completion, we use the KronEM algorithm~\cite{kronem}, which is built upon the Kronecker generative graph model and is the current state-of-the-art algorithm in the literature. Based on an observable network $G$, KronEM estimates the parameter matrix $\Theta$ used to generate the full network $\Theta^K$ representing the $K$th Kronecker power of $\Theta$, where $K$ is a positive integer such that $N_0^{K-1}<N+M\le N_0^K$. Let $({\bf A}_R^{(M)},\sigma)$ denote a permutation matrix, where $\sigma$ indicates a permutation of the set $\{1,\cdots,N+M\}$ and $\sigma(u)$ is the index of node $u$ in the graph $R^{(M)}$ after permutation. The first $N$ elements of $\sigma$ map the nodes in $G$ while the remaining $M$ elements map the nodes in the missing part. Then, the likelihood $\mathbb{P}({\bf A}_R^{(M)},\sigma|\bm{\Theta})$ can be expressed as
\begin{equation}\label{eq:2}
\mathbb{P}({\bf A}_R^{(M)},\sigma|\bm{\Theta}) = \displaystyle\prod_{a_{uv} =1}[\bm{\Theta}^K]_{\sigma(u)\sigma(v)} \displaystyle\prod_{a_{uv} = 0}(1-[\bm{\Theta}^K]_{\sigma(u)\sigma(v)}), \nonumber
\end{equation}
where $a_{uv}$ denotes the $(u,v)$th element of the matrix ${\bf{A}}_R^{(M)}$, and $[\bm{\Theta}^K]_{\sigma(u)\sigma(v)}$ denotes the $(\sigma(u),\sigma(v))$th element of the matrix $\bm{\Theta}^K$. 
As the matrix $\bm{\Theta}^K$ is a probabilistic representation of ${\bf A}_R^{(M)}$, we also obtain the  missing parts ${\bf Z}_1$ and ${\bf Z}_2$ by assigning the value of every entry in ${\bf Z}_1$ and ${\bf Z}_2$ to be zero or one according to a series of {\em Bernoulli coin-tosses} with the mapped entries in $\bm{\Theta}^K$ as the probabilities. The detailed steps will be discussed in Section~\ref{sec:4}.

\subsection{Influential Node Selection by Ranking}\label{sec:3c}
Network completion can be seen as a statistical learning process, as we predict the value of entries in the missing part of the network by leveraging information in the observed part. Thus, after obtaining the missing part via inference, we note that using the whole recovered nodes may lead to an inaccurate detection of communities due to two types of {\em errors}. One type stems from the prediction model, while the other stems from random errors that occur during the Bernoulli series used to project a probabilistic value to zero or one. While the prediction error can be reducible, the random error is irreducible. For this reason, the more recovered nodes we include, the higher the sum of errors. On the other hand, using just a very small portion of missing nodes is unlikely to provide correct community structures, since there is insufficient {\em information} available to the community detection model.

Since our eventual goal is to recover the true community structures, it is intuitive to add only nodes that are useful in the community detection process. To assess the usefulness of nodes to a social network, we rely on the concept of centrality and adopt the {\em degree centrality}, which measures the number of immediate neighbors of a node. Given an undirected graph, the degree centrality of node $u$, denoted by $\text{Cen}(u)$, is defined as the number of connections of a node (i.e., the number of incident edges of a node), and is computed as
\begin{equation}\label{eq:degreecen}
\text{Cen}(u) = \sum_{v=1}^{N+M} a_{uv}.
\end{equation} 

To select a subset of important nodes to recover, we rank the inferred nodes by first calculating their centrality measures and then sorting them in order of descending centrality. In the following, we formally define the concept of {\em influential nodes}.

\begin{definition}\label{def:2}
Let $\text{Cen}(u)$ denote a centrality measure of node $u \in (V \cup V_M)$. Then, $u$ is defined as an influential node if $u \notin V$ and $\text{Cen}(u) \geq \epsilon$, where $\epsilon > 0$ is a predefined threshold, $V$ is the set of observable nodes, and $V_M$ is the set of missing nodes. Here, $H \in \{0,1,\cdots,M\}$ denotes the cardinality of the set of influential nodes.
\end{definition}

The threshold $\epsilon$ signifies when the amount of acquired information outweighs the incurred errors from recovering parts of the true network, which means that recovering more than $H$ nodes can be harmful to the community detection process.

\subsection{Community Detection via Regularized NMF}\label{sec:3d}

Although there are several recent attempts to detect communities based on deep learning (e.g., link embedding~\cite{linkblack} and node embedding~\cite{vgraph}), matrix factorization-based approaches are still commonly used tools in the detection of overlapping communities in social networks~\cite{bigclam,BNMF,newnmf}. The primary benefit of these approaches lies in their {\em scalability} since many efficient techniques for solving NMF problems have been developed~\cite{nmfsolve}---the performance superiority of NMF-aided community detection will be empirically shown  later (refer to Table~\ref{table:combineperf}). Additionally, the NMF-based approaches aim at detecting community structures in the whole given network in a deterministic manner, which often requires less effort to find the optimize parameter setting than those based on the label propagation. In this subsection, we describe how community detection can be transformed into a regularized NMF-aided optimization problems and elaborate on detailed steps for solving the combined problem of graph inference and community detection.

From the previous steps, recall that we have an observation matrix ${\bf A}$, a list of influential nodes, and their corresponding ranking by the degree centrality measure. Furthermore, all recovered matrices ${\bf A}^{(i)}_R$ for $i \in \{0,1,\cdots,H\}$ are available. Due to the fact that $\mathbb{P}(u,v) = 1 - \exp(-{\bf F}_{u} {\bf F}^{\top}_{v})$ according to the AGM, the likelihood $\mathbb{P}({\bf A}_R^{(i)}|{\bf F})$ in (\ref{eq:1}) can be rewritten as
\begin{equation}\label{eq:5}
\mathbb{P}({\bf A}_R^{(i)}|{\bf F}) = \prod_{a_{uv}^{(i)} =1}(1 - \exp(-{\bf F}_u{\bf F}_v^\top)) \prod_{a_{uv}^{(i)} = 0}(\exp(-{\bf F}_u{\bf F}_v^\top)),\nonumber
\end{equation}
where $a_{uv}^{(i)}$ denotes the $(u,v)$th element of the matrix ${\bf{A}}_R^{(i)}$. Thus, we have
\begin{equation}\label{eq:6}
\log(\mathbb{P}({\bf A}_R^{(i)}|{\bf F})) = \sum_{a_{uv}^{(i)} =1}\log(1 - \exp(-{\bf F}_u{\bf F}_v^\top)) - \sum_{a_{uv}^{(i)} =0}({\bf F}_u{\bf F}_v^\top).
\end{equation}
Suppose that $f(\textbf{X}) = 1 - \exp(-(\textbf{X}))$ for a matrix $\textbf{X}$. Then, we obtain a matrix $f({\bf F}{\bf F}^\top)$ that probabilistically approximates the adjacency matrix ${\bf A}^{(i)}_R$. To estimate the difference between two matrices ${\bf A}^{(i)}_R$ and $f({\bf F}{\bf F}^\top)$, instead of using the Euclidean distance metric, we utilize the negative log-likelihood in~(\ref{eq:6}) as a loss function $\mathcal{D}$, which indicates that 

\begin{equation}\label{eq:D1}
\mathcal{D}({\bf A}^{(i)}_R, f({\bf F}{\bf F}^\top)) = -\log(\mathbb{P}({\bf A}_R^{(i)}|{\bf F})). 
\end{equation}

As a result, the optimization problem in~(\ref{eq:1}) can then be cast into a regularized NMF formulation as
\begin{equation}\label{eq:7}
(\hat{{\bf F}}, \hat{i})=\argmin_{{\bf F} \geq 0, i \in\{0,1,\cdots,H\} } \mathcal{D}({\bf A}^{(i)}_R, f({\bf F}{\bf F}^\top))-\lambda\log(i+1),
\end{equation}
where the objective function in~(\ref{eq:7}) is referred to as the {\em regularized loss} in our setup.

\section{Proposed \textsf{KroMFac} Framework}\label{sec:4}
In this section, to provide a complete solution to the problem of community detection in a partially observable graph, we present \textsf{KroMFac}, a novel framework that consists of the following three major phases: 1) network completion, 2) node ranking and selection, and 3) community detection. The overall procedure is described in Algorithm~\ref{al1}. The observable graph $G$, the number of missing nodes, $M$, and the number of communities, $C$, are the key input  parameters of the algorithm. The dimension of the parameter matrix $\bm{\Theta}$ is given by $N_0 \times N_0$, and we initialize $\bm{\Theta}$ as a randomly generated matrix $\bm{\Theta}_{\text{init}} \in [0,1]^{N_0 \times N_0}$. Further parameters serve as control parameters. In particular, $\epsilon$ plays a central role in determining the set of influential nodes and are introduced in detail in Section~\ref{sec:3c}, and $\lambda$ controls the impact of regularization, which can be quantified via an empirical study. The parameter $\delta > 0$ serves as a threshold to decide to which communities each node belongs (i.e., the degree of membership of nodes), and can be estimated for a given network~\cite{bigclam}. Finally, the parameter $\eta_{\text{detect}}$ is an arbitrarily small positive constant used as stopping criterion during community detection (i.e., convergence criteria). As the output of Algorithm~\ref{al1}, we define $\psi$ as the set of detected communities.

We assume that all communities initially have no members. To find the true community structures of the incomplete input graph, we first fully recover the graph via the function \textsf{GraphRecv} (refer to Section~\ref{graphrecv}). By analyzing the recovered graph via the function \textsf{NodeSelect}, we then select $H$ influential nodes and determine the ranking vector ${\bm r \in \N^H}$ that represents the indices of ranked influential nodes and plays a crucial role in community detection accuracy (refer to Section~\ref{noderank}). In this step, specifically, we solve the joint optimization problem described in~(\ref{eq:7}) through exhaustive search over $i$ by sequentially connecting influential nodes to the existing observable graph based on the order in ${\bm r}$. For each $i$, we acquire a corresponded graph $R^{(i)}$ and its adjacency matrix ${\bf A}^{(i)}_R$ (see Definition~\ref{def:1}). By using the function \textsf{CommunDet} implemented via the state-of-the-art NMF-based community detection method, we are then capable of obtaining the loss function $\mathcal{D}$ associated with affiliation matrix ${\bf F}$ given the input matrix ${\bf A}^{(i)}_R$. As a result, we obtain $\mathcal{D}_{\min}$ as the smallest value of $\mathcal{D}-\lambda\log(i+1)$ (i.e., the regularized loss), which in turn provides us with the optimal $\hat{{\bf F}}$ and $\hat{i}$. Every entry $\hat{{\bf F}}_{uc}$ in the optimal affiliation matrix $\hat{{\bf F}}$ then describes the likelihood that the node $u \in (V \cup V_M)$ belongs to community $c \in\{1,\cdots,C\}$. Therefore, it is possible to recover the community structures $\psi$ by assigning node $u$ to community $c$ if the corresponding entry $\hat{{\bf F}}_{uc}$ is greater than or equal to the threshold $\delta$. In the following, we elaborate on each major phase of the \textsf{KroMFac} framework.
\begin{algorithm}[t]
\DontPrintSemicolon
\KwIn{$G, M, C,N_0,\bm{\Theta}_{\text{init}},\epsilon,\delta, \eta_{\text{detect}}, \lambda$}
\KwOut{$\psi$}
\textbf{Initialization}: $ \mathcal{D}_{\min}\leftarrow \infty$; ${\psi}[c] \leftarrow \{\emptyset\}$ for $c \in \{1,\cdots,C\}$\\
\SetKwBlock{Begin}{function}{end function}
\Begin($\textsf{KroMFac}$)
{
  ${\bf A} \leftarrow$ Adjacency matrix of $G$\;
  ${\bf A}^{(M)}_R \leftarrow$ \textsf{GraphRecv}(${\bf A},N_0,\bm{\Theta}_{\text{init}},M$)\;
  $(H,{\bm r}) \leftarrow$ \textsf{NodeSelect}(${\bf A}^{(M)}_R,M,\epsilon$)\;
  \For{$i$ {\upshape \bf from} $1$ {\upshape \bf to} $H$}
  {
  $R^{(i)} \leftarrow$ Connect nodes \{${\bm r}[1],\cdots,{\bm r}[i]$\} to $G$\;
    ${\bf A}^{(i)}_R \leftarrow$ Adjacency matrix of $R^{(i)}$\;
    $(\mathcal{D},{\bf F}) \leftarrow$ \textsf{CommunDet}(${\bf A}^{(i)}_R, C, \eta_{\text{detect}}$)\;
    $\mathcal{D} \leftarrow \mathcal{D} - \lambda\log(i+1)$\;
    \If {$\mathcal{D}_{\min} < \mathcal{D}$}
    {
      $\hat{{\bf F}} \leftarrow {\bf F}$\;
      $\hat{i} \leftarrow i$\;
      $\mathcal{D}_{\min} \leftarrow \mathcal{D}$\;
      }
  }
  \For {$u$ {\upshape \bf from} $1$ {\upshape \bf to} $N+M$} 
  {
  	\For{$c$ {\upshape \bf from} $1$ {\upshape \bf to} $C$} {
		\If {$\hat{{\bf F}}_{uc} \geq \delta$} {
			${\psi}[c] \leftarrow {\psi}[c] \cup \{u\} $\;
		}
	}
  }
  \Return{${\psi}$}
}
\caption{\textsf{KroMFac}}\label{al1}

\end{algorithm}

\subsection{Network Completion}\label{graphrecv}
The first step of our framework is the inference of missing parts of the graph from priors on the observable matrix ${\bf A}$ and the number of missing nodes, $M$, by using the function \textsf{GraphRecv} implemented via the KronEM algorithm. First, from an initialized $\bm{\Theta}_{\text{init}}$, the E-step samples the missing parts ${\bf Z}_1$ and ${\bf Z}_2$, and the permutation $\sigma$. In the M-step, a stochastic gradient descent process subsequently optimizes the parameter matrix $\bm{\Theta}$ given the samples obtained in the E-step. The EM iteration alternates between performing the E-step and M-step according to the following expressions, respectively:

\textit{E-step}:
\begin{equation*}
\left({\bf Z}_1^{(t)},{\bf Z}_2^{(t)},\sigma^{(t)}\right) \sim \mathbb{P}({\bf Z}_1,{\bf Z}_2,\sigma|{\bf A}, \bm{\Theta}^{(t)}),
\end{equation*}

\begin{figure*}[t]
    \begin{center}
            \includegraphics[width=\linewidth]{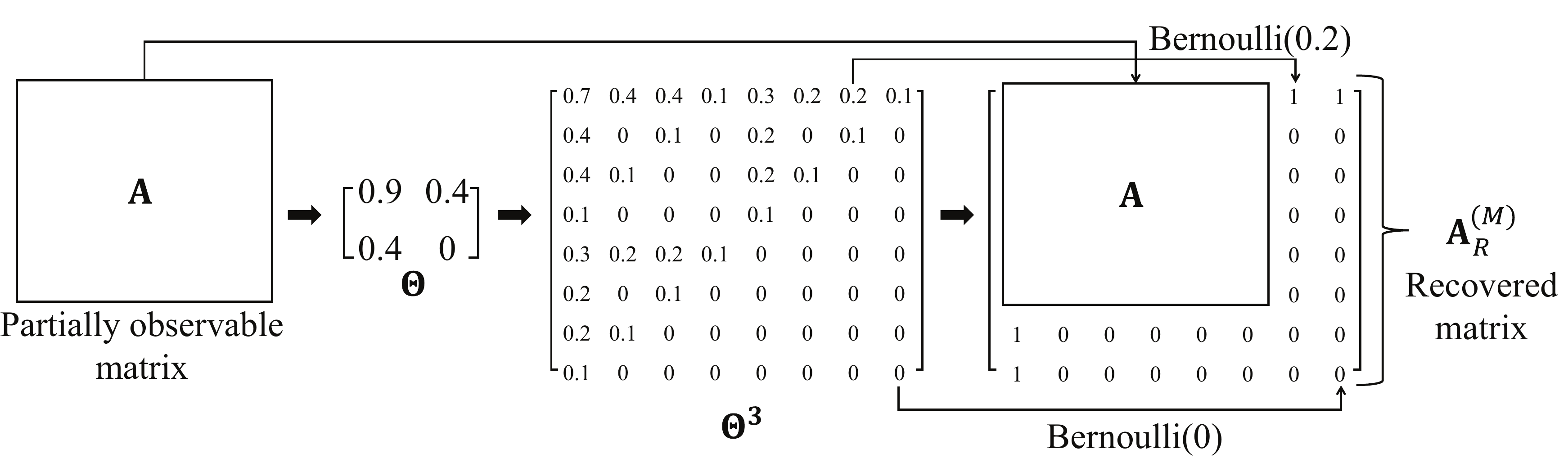}
            \caption{{An illustration of the recovery phase in our \textsf{KroMFac} framework. Here, parameters are set to the following values: $N = 6, M = 2, N_0 = 2, K = 3$, and $\sigma(u) = u$ for $u \in (V \cup V_{M})$.}}
            \label{fig:graphrecv}
    \end{center}
\vspace{-0.1in}
\end{figure*}

\textit{M-step}:
\begin{equation*}
\bm{\Theta}^{(t+1)} = \argmax_{\bm{\Theta} \in (0,1)^{N_0}} \mathbb{E}[\mathbb{P}({\bf Z}_1^{(t)},{\bf Z}_2^{(t)},\sigma^{(t)},{\bf A}| \bm{\Theta})], 
\end{equation*}
where the superscript $(t)$ denotes the iteration index.
Then, we generate the stochastic adjacency matrix $\bm{\Theta}^K$. To create the fully recovered matrix ${\bf A}^{(M)}_R$, for the first $N$ rows and columns of ${\bf A}_R^{(M)}$, we replicate the entries of matrix ${\bf A}$ in the upper left of matrix ${\bf A}^{(M)}_R$. To infer the missing part (i.e., the last $M$ rows and columns of ${\bf A}_R^{(M)}$), we consecutively run the {\em Bernoulli trials} with the probability $\bm{\Theta}^K_{\sigma(u)\sigma(v)}$ and then map the value of the missing entry in row $u$ and column $v$ to \textit{one} if a success occurs and {\em zero} otherwise. Since the adjacency matrix ${\bf A}_R^{(M)}$ is symmetric, we only need to repeat this process $MN + \frac{M^2}{2}$ times. An example of this network completion phase is illustrated in Figure~\ref{fig:graphrecv} when $N = 6, M = 2, N_0 = 2, K = 3$, and $\sigma(u) = u$ for $u \in (V \cup V_{M})$. In this example, since the last two rows and columns of $\bm{\Theta}^3$ correspond to two recovered nodes, we execute Bernoulli trials on each non-zero entry in this part to obtain the recovered matrix ${\bf A}_R^{(M)}$. If the number of missing edges can be estimated (see~\cite{kronem} for details), then the termination of this final step can be accelerated for sparse graphs since the Bernoulli trials can be terminated once the number of entries with a value that is equal to \textit{one} exceeds the predicted number of edges.
\subsection{Node Ranking and Selection}\label{noderank}
\begin{figure*}[t]
    \begin{center}
            \includegraphics[width=\linewidth]{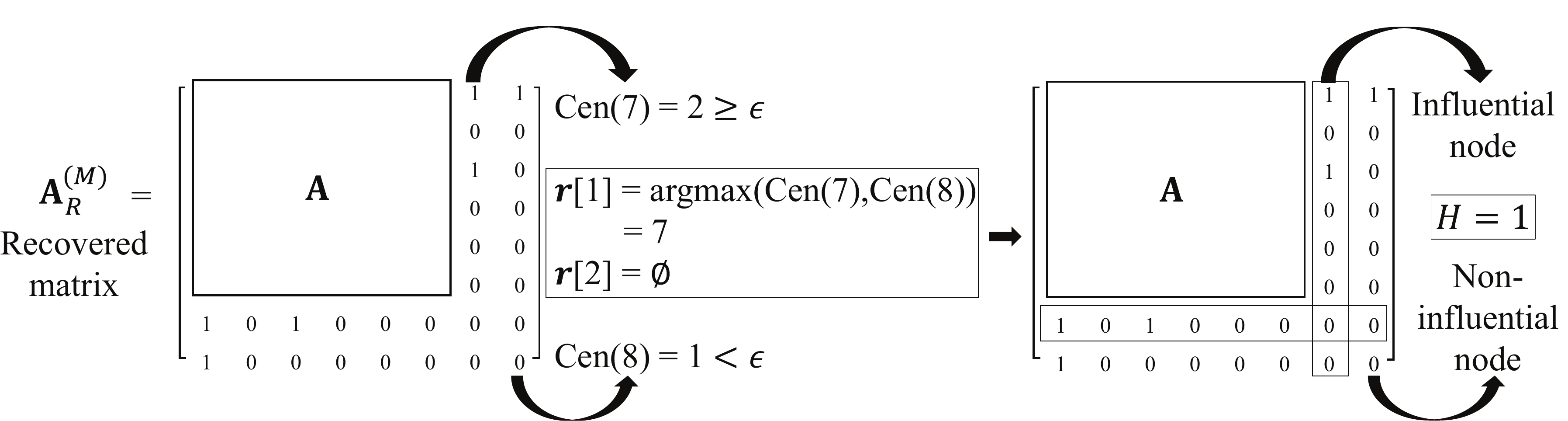}
            \caption{An illustration of the node ranking and selection phase. Here, parameters are set to the following values: $\epsilon = 2$ and $M = 2$.}
            \label{fig:nodeselect}
    \end{center}
\vspace{-0.1in}
\end{figure*}

Based on the output of the \textsf{GraphRecv} algorithm (i.e., the recovered matrix ${\bf A}^{(M)}_R$), the function \textsf{NodeSelect} ranks missing nodes and then selects influential nodes from the set of ranked candidates. In our work, we focus on the well-known centrality measure for ranking nodes, namely degree centrality $\text{Cen}(u)$ for node $u \in (V \cup V_M)$. 
After computing the degree centrality of missing nodes as in (\ref{eq:degreecen}), we introduce the ranking vector ${\bm r} \in \N^M$ to record their centrality ranking. Since we aim to select nodes whose centrality measures are greater than a given threshold $\epsilon$ (refer to Section~\ref{sec:3c}), only $H \leq M$ most influential nodes are associated with ${\bm r}$. For example, ${\bm r}[u]$ represents the index of the node ranked at the $u$th position in the list. The function \textsf{NodeSelect} returns the number of influential nodes, $H$, and the ranking vector ${\bm r \in \N^H}$. Figure~\ref{fig:nodeselect} shows a simple illustration of this node ranking and selection phase when $\epsilon = 2$ and $M = 2$. In this figure, due to the fact that the last two rows and columns in the input matrix ${\bf A}_R^{(M)}$ correspond to two recovered nodes, we calculate the degree centrality of the two nodes and select the seventh placed node as an influential node since its centrality is greater than or equal to $\epsilon$.

{\bf Discussion}. Conceptually, numerous centrality measures (e.g., eigenvector centrality and Katz centrality) can also be applied to obtain the ranking of nodes, but it turns out that they do not lead to better performance since the number of immediate neighbors of a node (rather than the number of higher-order neighbors) plays a crucial role in determining the performance of community detection under our setting. As an example, according to a prior study \cite{katzweakness}, nodes with a high Katz centrality may act as a bridge connecting different communities. In such a case, losing these nodes would typically not affect the performance of community detection negatively. More importantly, the NMF-based objective function in (\ref{eq:7}) directly optimizes the matrix ${\bf F}$ based on the observation from direct edges in the matrix ${\bf A}_R^{(i)}$. In this context, the degree centrality is a suitable measure to identify influential nodes that greatly alter ${\bf F}$. We empirically validate this claim in the appendix by comparing the performance along with two centrality measures.

\subsection{Community Detection}

To solve the problem of community detection, we adopt the state-of-the-art NMF-based detection algorithm, named BIGCLAM ~\cite{bigclam}, via as the function \textsf{CommunDet}, which can however be replaced by other community detection methods that return an affiliation matrix. For given $i$, we solve (\ref{eq:7}) using a block coordinate gradient ascent algorithm~\cite{nmfsolve}. The optimization process is terminated when the change in each iteration, denoted by $\Delta\mathcal{D} > 0$, is less than an arbitrarily small threshold $\eta_{\text{detect}} > 0$. The algorithm returns the loss function $\mathcal{D}$ and the corresponding matrix ${\bf F}$.

\subsection{Analysis of  Computational Complexity}
In this subsection, we analyze the computational complexity of the \textsf{KroMFac} framework. Since our framework consists of three major phases including network recovery, influential node selection, and community detection, we elaborate on the complexity analysis of each phase. To reduce the complexity, we take advantage of the property that real-world social networks usually have a sparse and low-rank matrix structure~\cite{Richard}.
The network completion phase is based on the KronEM algorithm, which was shown in~\cite{kronem} to have the complexity of $\mathcal{O}(|E|\log |E|)$, where $|E|$ is the number of edges in the partially observable network $G$. In the node selection phase, the computation of degree centrality dominates the complexity, which is given by $\mathcal{O}(|E|)$ in sparse graphs. In the community detection phase, an almost linear complexity in $N$ can be achieved via the approach in~\cite{bigclam}, where $N$ denotes the number of observable nodes. Here, while the community detection process is repeated $H$ times, one can see that $H \ll M$ (see Section~\ref{syndata}), where $H$ and $M$ denote the numbers of influential nodes and missing nodes, respectively. From the fact that $M$ is smaller than $N$, we can deduce that the complexity of this phase is bounded by $\mathcal{O}(N)$. Hence, the total computational complexity of \textsf{KroMFac} is finally given by $\mathcal{O}(|E|\log|E|)$.

\section{Experimental Evaluation}\label{sec:5}

In this section, we first describe both synthetic and real-world datasets. We also present two baseline schemes for community detection as a comparison, which corresponds to our ablation study during each step of the proposed \textsf{KroMFac} framework. By adopting the NMI as a popular information-theoretic performance metric, we then present the performance of our community detection framework and compare it against the two baseline schemes. 

\subsection{Datasets}\label{sec:5a}
To evaluate the community detection performance of our approach, we rely on datasets for which ground-truth communities are explicitly labeled. In the following, both synthetic and real-world datasets across various domains are taken into account.
\subsubsection{Synthetic Datasets}\label{sec:5a1}
\begin{table}[t]
\centering
\caption{LFR parameters of 18 synthetic graphs. Here, M and k denote $10^6$ and $10^3$, respectively. NN: the number of nodes, AD: average degree, MD: maximum degree, MinC: minimum community size, MaxC: maximum community size, DE: degree exponent, CSE: community size exponent, MP: mixing parameter, ON: the number of overlapping nodes, CMN: the number of communities per node.}
\label{table:1}
\begin{tabular}{|c|c|c|c|c|c|c|c|c|c|c|c|c|c|}
\hline
{\textbf{Graphs}} & \textbf{NN} & \textbf{AD} & \textbf{MD} & \textbf{MinC} &\textbf{MaxC} &\textbf{DE} &\textbf{CSE} &\textbf{MP} &\textbf{ON} &\textbf{CMN} \\ \hline
Graph 1          & 10k    & 10    & 50  & 10    & 50    & 3 
& 1    & 0.2    & 500 & 30    \\ 
Graph 2          & 10k    & 10    & 50  & 10    & 50    & 3 
& 1    & 0.25    & 500 & 30    \\ 
Graph 3          & 10k    & 10    & 50  & 10    & 50    & 3 
& 1    & 0.28    & 500 & 30    \\ 
Graph 4          & 10k    & 10    & 50  & 10    & 50    & 3 
& 1    & 0.2    & 500 & 3    \\ 
Graph 5          & 10k    & 10    & 50  & 10    & 50    & 3 
& 1    & 0.25    & 500 & 3    \\ 
Graph 6          & 10k    & 10    & 50  & 10    & 50    & 3 
& 1    & 0.28    & 500 & 3    \\ 
Graph 7          & 100k    & 10    & 50  & 10    & 500    & 3 
& 1    & 0.2    & 5000 & 30    \\ 
Graph 8          & 100k    & 10    & 50  & 10    & 500    & 3 
& 1    & 0.25    & 5000 & 30    \\ 
Graph 9          & 100k    & 10    & 50  & 10    & 500    & 3 
& 1    & 0.28    & 5000 & 30    \\ 
Graph 10          & 100k     & 10    & 50 & 10    & 500    & 3 
& 1    & 0.2    & 5000 & 3    \\ 
Graph 11          & 100k     & 10    & 50  & 10    & 500    & 3 
& 1    & 0.25    & 5000 & 3    \\ 
Graph 12          & 100k    & 10    & 50  & 10    & 500    & 3 
& 1    & 0.28    & 5000 & 3    \\ 
Graph 13          & 1M    & 10    & 50  & 10    & 5000    & 3 
& 1    & 0.2    & 50000 & 30    \\ 
Graph 14          & 1M    & 10    & 50  & 10    & 5000    & 3 
& 1    & 0.25    & 50000 & 30    \\ 
Graph 15          & 1M    & 10    & 50  & 10    & 5000    & 3 
& 1    & 0.28    & 50000 & 30    \\ 
Graph 16          & 1M     & 10    & 50  & 10    & 5000    & 3 
& 1    & 0.2    & 50000 & 3    \\ 
Graph 17          & 1M     & 10    & 50  & 10    & 5000    & 3 
& 1    & 0.25    & 50000 & 3    \\ 
Graph 18         & 1M     & 10    & 50  & 10    & 5000    & 3 
& 1    & 0.28    & 50000 & 3    \\ 
\hline 
\end{tabular}
\end{table}
We construct synthetic graphs via the extended Lancichinetti-Fortunato-Radicchi (LFR) benchmark~\cite{lfr}, which is built upon a generative model that creates nodes along with prior known community labels. The benchmark is capable of generating graphs that replicate important features of real social networks, such as the power-law degree distribution and overlapping communities. To create an LFR graph, ten parameters need to be specified, which are summarized in Table~\ref{table:1}. While parameters such as the number of nodes, average degree, maximum degree, maximum and minimum community size, and degree exponent are rather straightforward to understand, we explain the remaining four parameters:
\begin{itemize}
\item The community size exponent refers to the exponent parameter from a power-law approximation of the distribution of the number of nodes in communities.
\item The mixing parameter, denoted by $\mu$, controls the proportion of random edges to total edges. For example, if $\mu = 0.3$, then the LFR benchmark produces a graph such that approximately 70\% of edges link to nodes within the same community, while the remaining 30\% connect to nodes in other randomly selected communities. This parameter is sensitive to the performance of community detection. In general, if $\mu$ is closer to one, then the community structures become weaker. On the other hand, when $\mu$ is closer to zero, one can expect high detection performance since community structures can be easily identified.
\item The number of overlapping nodes refers to the number of nodes in the graph that belong to more than one community.
\item The number of communities per node indicates the average number of communities to which each of the overlapping nodes belongs. 
\end{itemize}
To cover various domains of network applications, we generate 18 LFR graphs with differing parameter settings according to~\cite{xie2013overlapping} as specified in Table~\ref{table:1}, where we show the various values chosen for representative parameters such as NN, MP, ON, and CMN.

\subsubsection{Real-World Datasets}\label{sec:5a2}

\begin{table}[t]
\centering
\caption{Statistics of the six real-world datasets. Here, M and k denote $10^6$ and $10^3$, respectively.}
\label{table:2}
\begin{tabular}{|c|c|c|c|c|c|}
\hline
\multicolumn{1}{|l|}{\bf Dataset} & \multicolumn{1}{l|}{\bf NN} & \multicolumn{1}{l|}{\bf NE} & \multicolumn{1}{l|}{\bf NC} & \multicolumn{1}{l|}{\bf ACS} & \multicolumn{1}{l|}{\bf CMN} \\ \hline
Amazon                             & 0.34M                  & 0.93M                  & 49k                   & 99.86                  & 14.83                  \\
DBLP                               & 0.43M                  & 1.3M                   & 2.5k                   & 429.79                 & 2.57    \\  
Youtube                           & 1.1M                & 3.0M                & 30k                 & 9.75                &0.26
\\
Facebook                             & 4k                  & 88k                  & 193                   & 22.93                  & 1.14                  \\
Twitter                               & 81k                  & 2.4M                   & 4k                   & 33.50                 & 1.65 \\  
Orkut                           & 3M                & 117M                & 6.3M                 & 34.86                &95.93
\\ \cline{1-6}            
\end{tabular}
\end{table}
To validate the applicability of our approach, six real-world datasets are also used for evaluation. More specifically, from the available SNAP datasets~\cite{snapdata} that have ground-truth communities, we use the Amazon product co-purchasing network~\cite{amazon}, the collaboration network of DBLP~\cite{dblp}, the Youtube video-sharing social network~\cite{youtube}, and the three friendship social networks of Facebook~\cite{facebook_data}, Twitter~\cite{facebook_data}, and Orkut~\cite{amazon}. The statistics of these datasets are summarized in Table~\ref{table:2}, and the basic characteristics of each network are described in the following:

\begin{itemize}
\item {\em The number of nodes (NN):} In the Amazon network, nodes represent products. In DBLP, nodes represent authors. In the Youtube, Facebook, Twitter, and Orkut networks, nodes represent users.
\item {\em The number of edges (NE):} In the Amazon network, edges connect products that are commonly purchased together. In DBLP, two authors are connected by an edge if they have co-authored a paper. In the Youtube, Facebook, Twitter, and Orkut networks, edges represent friendships between users.
\item {\em The number of communities (NC):} In the Amazon network, each product category corresponds to a ground-truth community. In DBLP, the publication venues are used as ground-truth communities. In the Youtube and Orkut networks, user-created groups are used as ground-truth communities. In the Facebook and Twitter networks, circles of users are used as ground-truth communities.
\item {\em Average community size (ACS):} The average number of nodes within communities.
\item {\em Community memberships per node (CMN):} The average number of communities that a node belongs to.
\end{itemize}

\subsection{Baseline Approaches (Ablation Study)}\label{sec:4b}

Due to the fact that community discovery in partially observable networks with both missing nodes and edges has never been studied in the literature, there is no state-of-the-art method that works appropriately under our network model. For this reason, we perform an ablation study by removing the network completion and/or influential node selection components from our \textsf{KroMFac} framework. To this end, we introduce two types of baseline schemes by taking into account some of the special cases of the proposed \textsf{KroMFac} framework.

\subsubsection{Baseline 1 (Community Detection)} 

As a na\"ive approach, the first baseline scheme (Baseline 1) aims to directly discover community structures based on an observable network via the NMF-aided detection method without recovering any nodes and edges. To this end, Baseline 1 solves an optimization problem such that the matrix ${\bf F}$  is found given an adjacency matrix ${\bf A}$ of the incomplete network. The problem formulation is thus given by
$\hat{{\bf F}}=\argmax_{{\bf F} \geq 0} \mathbb{P}({\bf A}|{\bf F} )$,
which corresponds to a special case with no regularization term where $i$ is set to zero in our joint optimization problem~(\ref{eq:1}). Similarly as in the methodology in Section~\ref{sec:3d}, the optimal $\hat{\bf F}$ can be easily acquired via the function \textsf{CommunDet} by replacing the input matrix ${\bf A}_R^{(i)}$ by ${\bf A}$. As \textsf{CommunDet} results in $\hat{{\bf F}}$ providing fuzzy information on the community memberships, we apply the hard-decision process (refer to lines 15--18 in Algorithm~\ref{al1}) to find the community structure $\psi$.

\subsubsection{Baseline 2 (Network Completion + Community Detection)}\label{sec5b2}

In addition to Baseline 1, to highlight the importance of node ranking and selection, we also present the second baseline scheme (Baseline 2) that performs community detection along with a full graph recovery. To this end, given an adjacency matrix ${\bf A}_R^{(M)}$ that is inferred by the network completion phase, Baseline 2 solves an optimization problem such that the matrix ${\bf F}$ is found. The problem formulation is thus given by
$\hat{{\bf F}}=\argmax_{{\bf F} \geq 0} \mathbb{P}({\bf A}^{(M)}_R|{\bf F} )$,
which corresponds to another special case with no regularization term, where $i$ is set to $M$  in~(\ref{eq:1}). Note that node ranking  is not necessary since all recovered nodes are inserted into the graph. To solve this problem, we first follow the steps similar to those in Section~\ref{sec:3b} to recover the matrix ${\bf A}^{(M)}_R$. After the network completion phase via the function \textsf{GraphRecv}, the optimal ${\bf F}$ can be acquired via \textsf{CommunDet} by assuming that $i=M$. Then, the hard-decision process  in Algorithm~\ref{al1} is performed to produce the final result $\psi$. 

Additionally, as alternatives to BIGCLAM,  we consider the following three state-of-the-art algorithms for community detection:  COPRA~\cite{COPRA}, SLPA~\cite{SLPA}, and vGraph~\cite{vgraph}, whose source codes are publicly available. COPRA enables each node to update its community assignment coefficients by allowing a node to have multiple labels; SLPA allows nodes to exchange labels according to pairwise interaction rules; and vGraph presents a generative model for joint community detection and node representation learning.\footnote{Instead of exploiting the number of communities as side information, COPRA and SLPA take advantage of the number of communities per node and the minimum and maximum sizes of communities, respectively, to recover the community structures.} In this case, we first recover the matrix ${\bf A}^{(M)}_R$ via the function \textsf{GraphRecv} and then obtain $\psi$ via alternative algorithms. 


\subsection{Performance Metric}\label{sec:5b}

To assess the performance of our \textsf{KroMFac} framework and two baseline schemes, we need to quantify the degree of agreement between the ground-truth communities and the detected communities. In particular, given a set of true labels and the set of labels assigned by the resulting community detection, we need to find the similarity between them.
While there are various ways to estimate the similarity, the NMI is one of the most widely used evaluation measures for community detection problems~\cite{bigclam,BNMF}, and is formally defined as in the following.
\begin{definition}[NMI~\cite{nmioverlap}]\label{def:3}
Assume that the community assignments are ${x_i}$ and ${y_i}$, where $x_i$ and $y_i$ indicate the labels of vertex $i$ in the true community $\mathcal{X}$ and the predicted community $\mathcal{Y}$, respectively. When the labels $x$ and $y$ are the values of two random variables $X$ and $Y$, following a joint distribution $\mathbb{P}(x,y) = \mathbb{P}(X = x, Y = y)$, the NMI between $\mathcal{X}$ and $\mathcal{Y}$ is given by
$\text{NMI}(\mathcal{X},\mathcal{Y}) = 1 - \frac{1}{2}\left(\frac{H(X|Y)}{H(X)}+\frac{H(Y|X)}{H(Y)}\right)$,
where $H(X) = -\sum_x \mathbb{P}(x) \log \mathbb{P}(x)$ is the Shannon entropy of $X$ and $H(X|Y) = -\sum_{x,y} \mathbb{P}(x, y) \log \mathbb{P}(x|y)$ is the conditional entropy of $X$ given $Y$. 
\end{definition}

It is worth noting that other popular metrics such as modularity \cite{newman2006modularity} and conductance \cite{leung2009towards} cannot correctly evaluate the quality of detected communities in this setting. The reason for this is that community detection based on such metrics is conducted using only a distorted network structure born of partial observation, as if the distorted network is the underlying true one. For this reason, an evaluation score resulting from these metrics does not correctly reflect the ground-truth communities (i.e., labels).

\subsection{Experimental Results}\label{sec:5c}

To create partially observable networks $G$ from the underlying true graphs $G'$, we adopt two graph sampling strategies from~\cite{graphsampling}. The first strategy, called {\em random node (RN)} sampling, selects nodes uniformly at random to create a sample graph. The second one, {\em forest fire (FF)} sampling, starts by picking a seed node uniformly at random and adding it to a sample graph (referred to as burning). Then, FF sampling burns a fraction of the outgoing links with nodes attached to them. This process is recursively repeated for each neighbor that is burned until no new node is selected to be burned. Both sampling strategies are known not to be biased towards high degree nodes, while FF sampling is capable of preserving the degree distribution of the true graph~\cite{graphsampling}. 
We create partially observable networks consisting of 70\% nodes from the original synthetic and real-world datasets mentioned in Sections~\ref{sec:5a1} and~\ref{sec:5a2}, respectively, by performing two aforementioned graph sampling strategies. To be consistent in evaluating the performance of community detection in the incomplete graphs, we also perform node deletion such that the ground-truth community structures $\psi$ do not contain nodes removed from the original graphs. 

Our empirical study is basically designed to answer the following four key research questions.
\begin{itemize}
\item {\em Q1}. To what extent does the parameter $\lambda$ in regularization affect the performance?
\item {\em Q2}. How close is the optimal $\hat{i}$ that is the solution to~(\ref{eq:7}) to the value of $i$ that maximizes the NMI? 
\item {\em Q3}. How does the performance change when different community detection methods are adopted?
\item {\em Q4}. How much does our \textsf{KroMFac} method enhance the performance over the baseline schemes?
\end{itemize}

\subsubsection{Sensitivity Analysis (Q1 \& Q2)}\label{syndata}
\begin{figure}[t!]
\begin{subfigure}[b]{.31\textwidth}
\centering
\includegraphics[width=4.1cm]{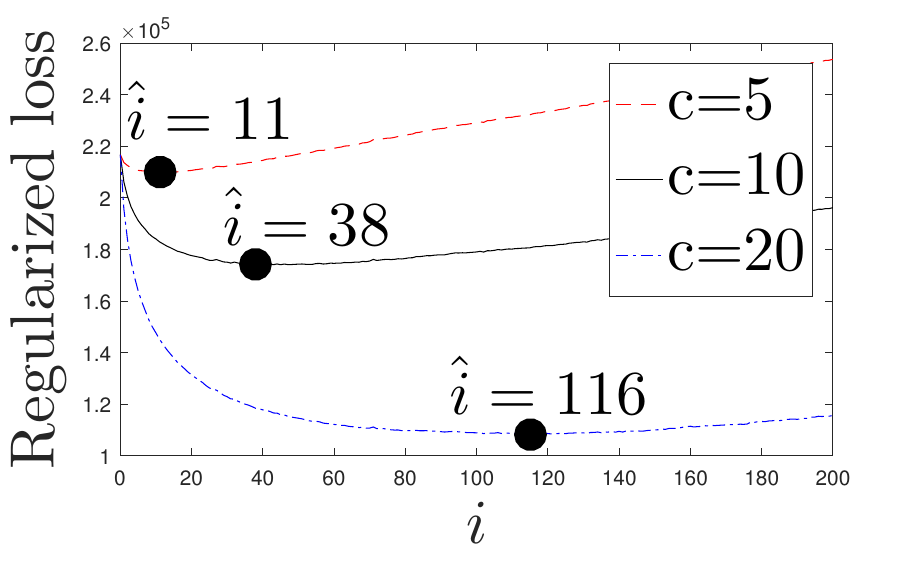}
\caption{Graph 1}\label{fig:sens1}
\end{subfigure}\hspace{1mm}
\begin{subfigure}[b]{.31\textwidth}
\centering
\includegraphics[width=4.1cm]{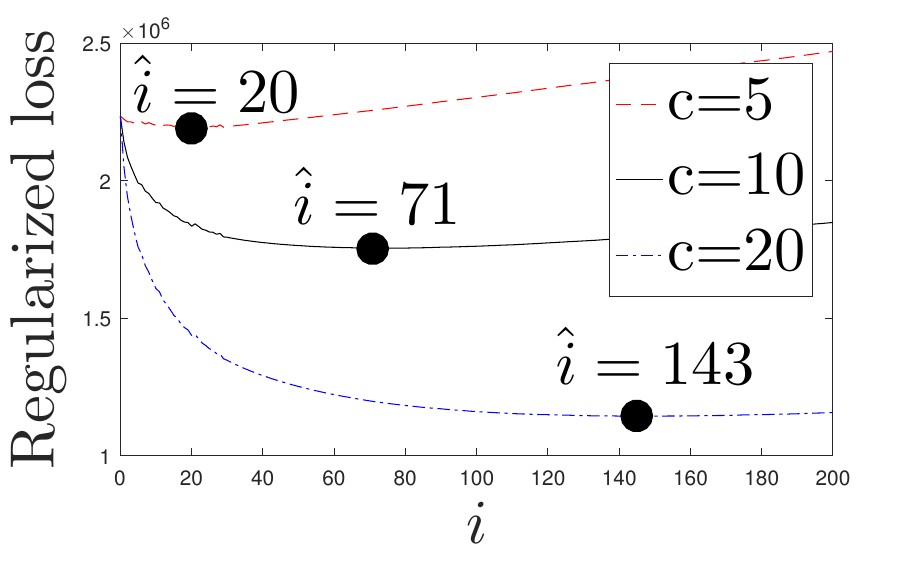}
\caption{Graph 7}\label{fig:sens7}
\end{subfigure}
\begin{subfigure}[b]{.31\textwidth}
\centering
\includegraphics[width=4.1cm]{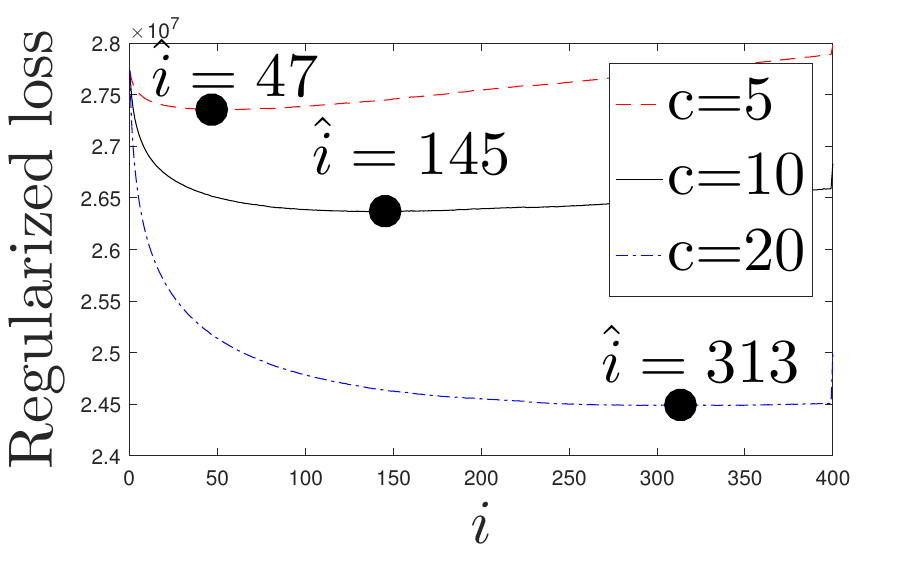}
\caption{Graph 13}\label{fig:sens13}
\end{subfigure}\hspace{1mm}
\caption{Regularized loss over the number of added nodes, $i$, according to different values of $c$. Here, black circles depict the points at which the minimum losses are attained.} 
\label{fig:sens}
\end{figure}
\begin{figure}[tbh!]
\begin{subfigure}[b]{.31\textwidth}
\centering
\includegraphics[width=4.3cm]{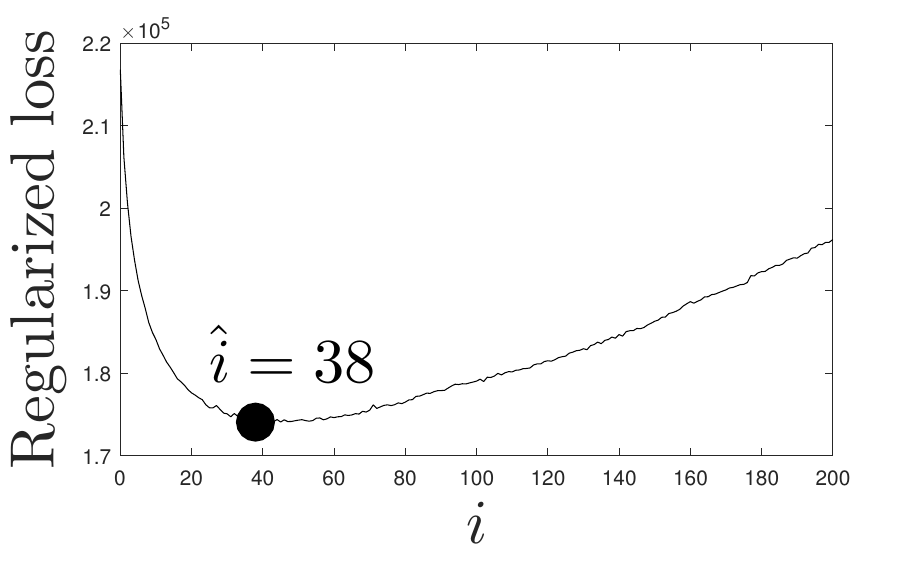}\hspace{1mm}
\includegraphics[width=4.3cm]{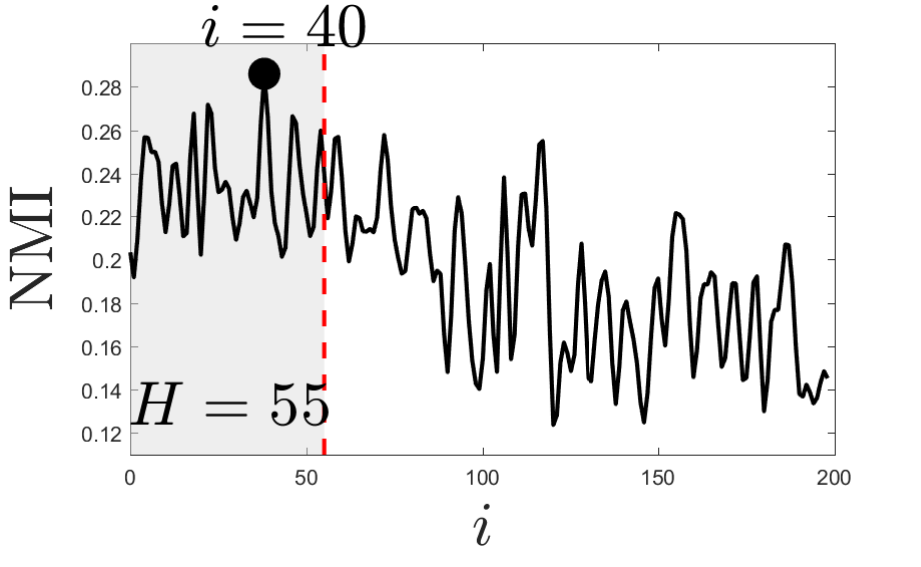}
\caption{Graph 1}\label{fig:6anew}
\end{subfigure}
\begin{subfigure}[b]{.31\textwidth}
\centering
\includegraphics[width=4.3cm]{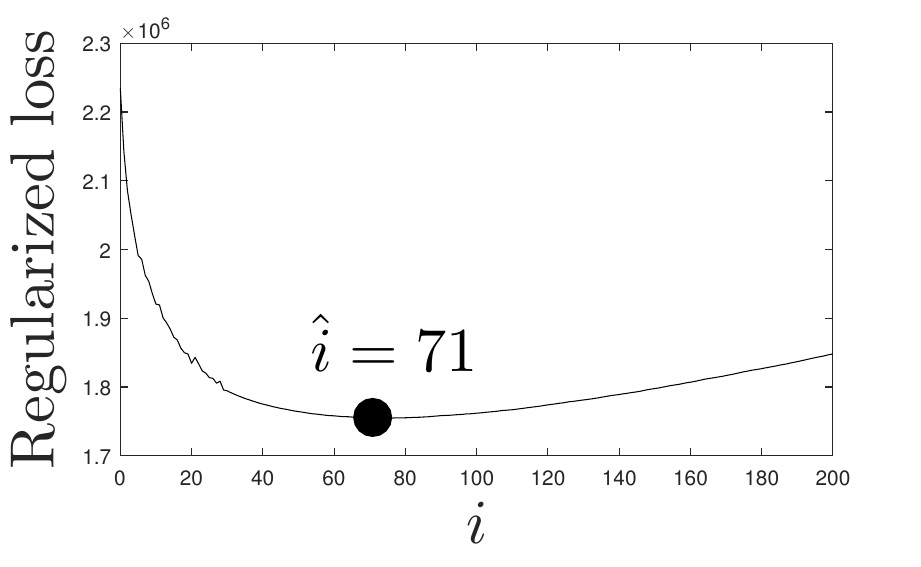}\hspace{1mm}
\includegraphics[width=4.3cm]{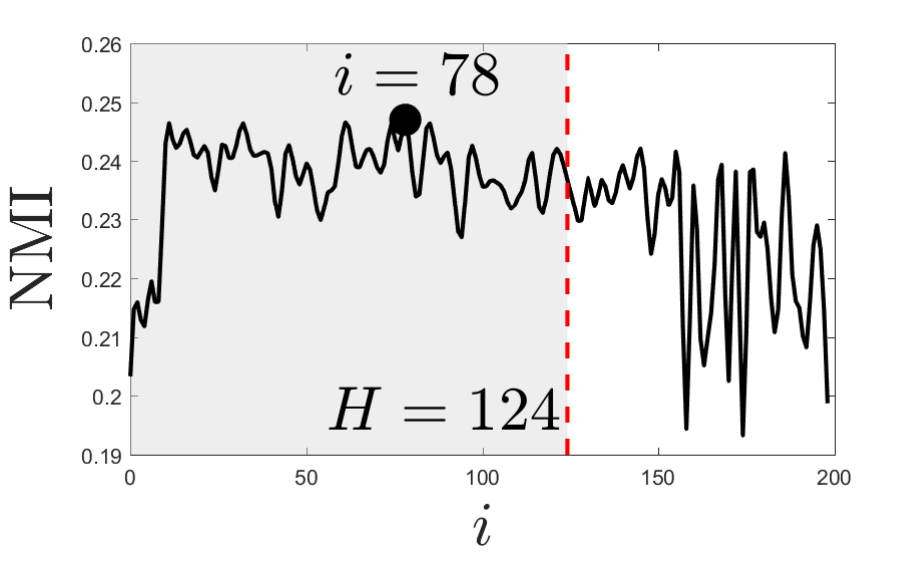}
\caption{Graph 7}\label{fig:6bnew}
\end{subfigure}
\begin{subfigure}[b]{.31\textwidth}
\centering
\includegraphics[width=4.3cm]{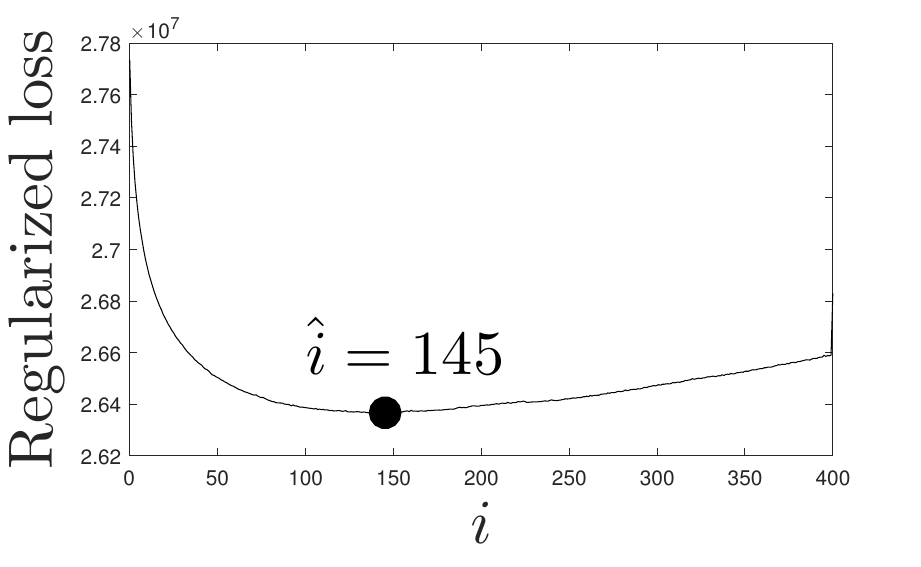}\hspace{1mm}
\includegraphics[width=4.3cm]{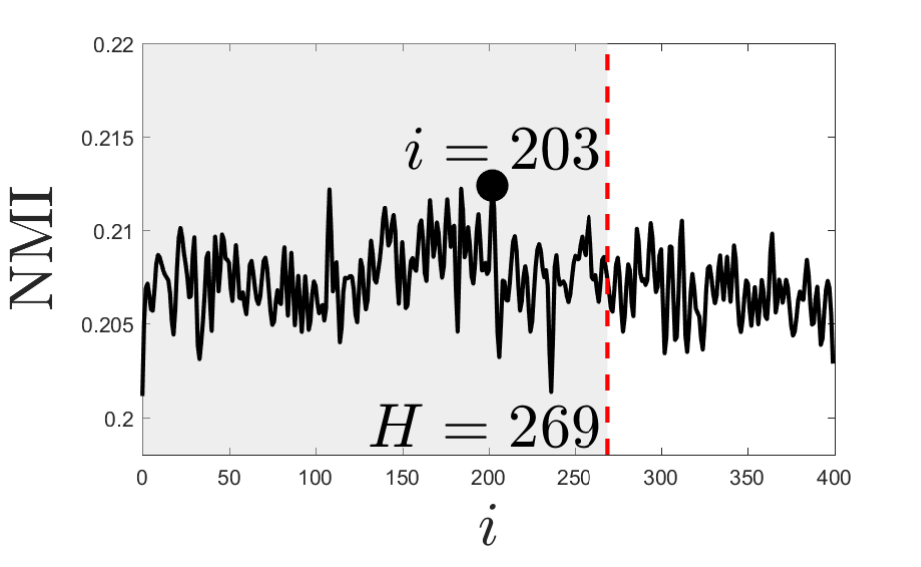}
\caption{Graph 13}\label{fig:6cnew}
\end{subfigure}
    \caption{Performance evaluation of \textsf{KroMFac} over $i$. Here, black circles depict the points (top) at which the minimum losses are attained and the points (bottom) at which the maximum NMIs are achieved. }
\label{fig:icomp}
\end{figure}
For the sensitivity analysis, we consider only the results where RN sampling is applied to Graphs 1, 7, and 13 of our synthetic graphs (each with different network sizes) since other cases follow similar trends.
First, we analyze the sensitivity of the parameter $\lambda$, which determines the impact of regularization in our optimization problem and plays a crucial role in determining performance of our \textsf{KroMFac} framework.  In Figure~\ref{fig:sens}, the regularized loss over the number of added influential nodes, $i$, is illustrated according to various values of $c>0$, where $\lambda=cN$. We find that setting $c$ to a small value (e.g., $c=5$) results in a value of $\hat{i}$ that is too low to compensate the loss function properly with the regularization term. In contrast, setting it to a high value (e.g., $c=20$) results in a value of $\hat{i}$ that falls out of the acceptable range of $i$,  i.e., $i > H$, due to over-regularization. We empirically verify that setting $\lambda = 10N$ manifests satisfactory performance in terms of regularized loss and NMI in the following experiments. We demonstrate that our setting of $\lambda$ is robust to the overall network attributes, including the network size. Furthermore, we investigate how close the optimal $\hat{i}$ that is the solution to~(\ref{eq:7}) is to the value of $i$ that maximizes the NMI. In Figures~\ref{fig:6anew}--\ref{fig:6cnew} (top), we illustrate the regularized losses (i.e., the objective function in (\ref{eq:7})) over $i$, where the minimum losses in Graphs 1, 7, and 13 are attained at $\hat{i}=38$, $\hat{i}=71$, and $\hat{i}=145$, respectively. In Figures~\ref{fig:6anew}--\ref{fig:6cnew} (bottom), we plot the NMI over $i$, where the maximum NMIs in Graphs 1, 7, and 13 are achieved at $i=40$, $i=78$, and  $i=203$, respectively. From these three graphs, we observe that adding more nodes and edges to the existing graph increases the NMI scores up to a certain number of nodes (e.g., $i=40$ in Figure~\ref{fig:6anew} (bottom)), but drops if more nodes are added due to a higher accumulated inference error, which verifies our assertion made in Section~\ref{sec:3}. The fact that the NMIs attained by the minimum losses are close to the maximum NMIs in Figures~\ref{fig:6anew}--\ref{fig:6cnew} is an indication that the solutions to~(\ref{eq:7}) also ensure satisfactory performance with regard to the NMI. Additionally, we empirically determine the threshold $\epsilon$, which plays a crucial role in specifying the number of influential nodes, $H$. When we set $\epsilon=\frac{k_{\max}}{2}$ as in~\cite{yuan2010fish}, the resulting value of $H$ in Graphs 1, 7, and 13 are 55,  124, and $269$, respectively, and the NMIs are computed by searching over $i\in\{1,\cdots,H\}$ (see the shaded areas in Figures \ref{fig:6anew}--\ref{fig:6cnew}), where $k_{\max}>0$ is the maximum degree of nodes in a recovered graph. 
 Since the threshold setting $\epsilon=\frac{k_{\max}}{2}$ leads to a reduction in computational complexity without loss of performance, we adopt it in our experiments in the following. This result also suggests that adding only a small number of nodes and their associated edges to the existing graph is sufficient to remarkably enhance the NMI performance.

\begin{table}[t]
\centering
\caption{NMI of Baseline 2 according to different community detection algorithms. Here, N/A indicates that the corresponding algorithm is unable to handle the computation due to memory consumption.}
\label{table:combineperf}
\bgroup
\def\arraystretch{1.2}
\resizebox{\columnwidth}{!}{%
\begin{tabular}{|c|c|c|c|c|}
\hline
\textbf{Name}
&\textbf{Baseline 2-BIGCLAM}
&\textbf{Baseline 2-COPRA}
&\textbf{Baseline 2-SLPA}
&\textbf{Baseline 2-vGraph} \\
 \hline
Graph 1                  &{\bf 0.1882}        & 0.0000 &0.0012  &0.0421                        \\
 \hline
Graph 2                  &{\bf 0.1710}        & 0.0000 &0.0020 &0.0315                    \\
 \hline
Graph 3                  &{\bf 0.1354}        & 0.0000 &0.0022  &0.0204                       \\
 \hline
Graph 4                 &{\bf 0.4014}        & 0.3024  &0.3121 & 0.2841                          \\
 \hline
Graph 5                 &{\bf 0.3245}        & 0.2201  &0.2031 & 0.2279                        \\
 \hline
Graph 6                  &{\bf 0.2648}       & 0.1704 &0.1425 & 0.1755                           \\
 \hline
DBLP                  &{\bf 0.1399}       & 0.1112 &0.1097 & N/A                         \\
 \hline
\end{tabular}
}
\egroup
\end{table}

\subsubsection{Comparison With Competing Community Detection Algorithms (Q3)}\label{Q2}
To see how the performance varies according to different community detection algorithms, we also evaluate the performance of COPRA~\cite{COPRA}, SLPA~\cite{SLPA}, and vGraph~\cite{vgraph} as specified in Section~\ref{sec5b2}. Table~\ref{table:combineperf} presents the NMI of Baseline 2 for fully recovered graphs, where RN sampling is applied to the DBLP network as well as Graphs 1–6, since the LFR parameters such as MP and CMN in synthetic datasets significantly affect the performance and the DBLP network is one of the datasets having the value of CMN comparable to that of Graphs 4--6 (refer to Table~\ref{table:2}). The results in Table \ref{table:combineperf} demonstrate the superiority of BIGCLAM over competing community detection methods. It is worth noting that all algorithms except for BIGCLAM almost entirely fail to recover the overlapping community structures for more difficult situations that have a higher value of CMN (e.g., Graphs 1--3). Surprisingly, despite its lower scalability,  the performance of vGraph is inferior to that of BIGCLAM in all experiments since vGraph does not perform appropriately for LFR graphs as in Graphs 1--6. This implies
 that a more sophisticated method may not provide better results for all types of network structures.

\subsubsection{Comparison With Two Baselines (Q4)}
Finally, the NMI of \textsf{KroMFac} and two baseline schemes for synthetic and real-world graphs are shown in Table~\ref{table:perfsyn}, where both RN and FF sampling strategies are applied and BIGCLAM is adopted for community detection. The NMI performance of BIGCLAM for {\em complete} networks without deleting nodes and edges is also shown to provide an upper bound of our approach. We find that our \textsf{KroMFac} framework noticeably outperforms the baselines for all synthetic and real-world datasets with substantial improvement rates of up to 22.85\% and 63.47\% over Baselines 1 and 2, respectively. Interestingly, our findings reveal that the NMI performance of \textsf{KroMFac} is not likely to be influenced by the network size; for example, similar  NMIs are obtained from Graphs 1, 7, and 13, whose network attributes are identical, except for the network size.
From the results, it is also clear that the performance of Baseline 2 is almost comparable to or even worse than that of Baseline 1, which shows that including the entirety of recovered nodes and edges is not beneficial. This result asserts the importance of the influential node selection component in our proposed framework.  
Furthermore, we observe that when the FF sampling strategy is applied, the performance of both \textsf{KroMFac} and two baselines is higher than the case of RN sampling; this implies that the improvement rate of \textsf{KroMFac} over baseline schemes is reduced for FF sampling. This is due to the fact that FF sampling tends to preserve the degree distribution of the true graph~\cite{graphsampling} and thus the community structures of the partially observable network would be less distorted than the case where RN sampling is used.


\begin{table}[t]
\centering
\caption{{NMI of \textsf{KroMFac} and two baseline schemes.}}
\label{table:perfsyn}
\bgroup
\def\arraystretch{1.2}
\resizebox{300pt}{!}{%
\begin{tabular}{|c|c|c|c|c|c|c|c|}
\cline{2-8} 
\multicolumn{1}{c}{\multirow{2}{*}{}} & \multicolumn{1}{|c|}{\multirow{2}{*}{\textbf{Name}}} & \multicolumn{1}{c|}{\multirow{1}{*}{\textbf{Complete}}} & \multicolumn{1}{c|}{\multirow{1}{*}{ {\bf \textsf{KroMFac}}}} & \multicolumn{1}{c|}{\multirow{1}{*}{\textbf{Baseline 1}}} & \multicolumn{1}{c|}{\multirow{1}{*}{\textbf{Baseline 2}}} & \multicolumn{2}{c|}{\textbf{Improvement rate (\%)}}                                                                     \\ \cline{7-8} 
\multicolumn{1}{c}{}                              &  \multicolumn{1}{|c}{}  & \multicolumn{1}{|c|}{\textbf{graph}}    &\multicolumn{1}{c|}{($X$)}                            & \multicolumn{1}{c|}{($Y$)}                                                 & \multicolumn{1}{c|}{($Z$)}  &\multicolumn{1}{c|}{$\frac{X-Y}{Y}\times100$} &\multicolumn{1}{c|}{$\frac{X-Z}{Z}\times100$} \\ \hline
\multicolumn{1}{|c|}{\multirow{36}{*}{\rotatebox{90}{\textbf{Synthetic datasets}}}} &Graph 1 (RN)                                            & \multirow{2}{*}{0.5125} & {\bf 0.2601}    & 0.2008 & 0.1882                                                      & {22.78}                                                & 27.64 \\
\multicolumn{1}{|c|}{}                              & Graph 1 (FF) &    & {\bf 0.3409 }   & 0.3226 & 0.2439                                                      & 5.37                                                & 28.46 \\
\cline{2-8} 
\multicolumn{1}{|c|}{}                              &Graph 2 (RN)                                            & \multirow{2}{*}{0.4234} & {\bf 0.1917}    & 0.1746 & 0.1710                                                      & 8.88                                                & 10.79 \\
\multicolumn{1}{|c|}{}                              & Graph 2 (FF) &    & {\bf 0.2667}    & 0.2510 & 0.2054                                                      & 5.90                                                & 23.00 \\
\cline{2-8} 
&Graph 3 (RN)                                            & \multirow{2}{*}{0.3886} & {\bf 0.1667}    & 0.1410 & 0.1354                                                      & 15.44                                                & 18.76 \\
\multicolumn{1}{|c|}{}                              & Graph 3 (FF) &    & {\bf 0.2566}    & 0.2234 & 0.1907                                                      & 12.94                                                & 25.70 \\
\cline{2-8} 
&Graph 4 (RN)                                            & \multirow{2}{*}{0.7825} & {\bf 0.4989}    & 0.4764 & 0.4014                                                      & 4.52                                                & 19.55 \\
& Graph 4 (FF) &    & {\bf 0.5182}    & 0.4782 & 0.4320                                                      & 7.71                                                & 16.62 \\
\cline{2-8} 
&Graph 5 (RN)                                            & \multirow{2}{*}{0.7214} & {\bf 0.3882}    & 0.3675 & 0.3245                                                      & 5.33                                                & 16.40 \\
& Graph 5 (FF) &    & {\bf 0.3968}    & 0.3517 & 0.3412                                                      & 11.36                                                & 14.00 \\
\cline{2-8} 
&Graph 6 (RN)                                            & \multirow{2}{*}{0.6821} & {\bf 0.3558}    & 0.3042 & 0.2648                                                      & 14.51                                                & 25.59 \\
& Graph 6 (FF) &    & {\bf 0.4153}    & 0.3204 & 0.28339                                                      & {\bf 22.85}                                                & 31.63 \\
\cline{2-8} 
&Graph 7 (RN)                                            & \multirow{2}{*}{0.4618} & {\bf 0.2266}    & 0.2033 & 0.1931                                                      & 5.70                                                & 28.48 \\
& Graph 7 (FF) &    & {\bf 0.3467 }   & 0.3269 & 0.2479                                                      & 10.28                                                & 14.78 \\
\cline{2-8} 
&Graph 8 (RN)                                            & \multirow{2}{*}{0.3654} & {\bf 0.1971}    & 0.1800 & 0.1537                                                      & 8.67                                                & 22.01 \\
& Graph 8 (FF) &    & {\bf  0.2630}    & 0.2304 & 0.2153                                                      & 12.39                                                &18.12 \\
\cline{2-8} 
&Graph 9 (RN)                                            & \multirow{2}{*}{0.3577} & {\bf 0.1709}    & 0.1473 & 0.1398                                                      & 13.79                                                & 18.20 \\
& Graph 9 (FF) &    & {\bf 0.2704}    & 0.2569 & 0.2036                                                      & 4.97                                                &24.72 \\
\cline{2-8} 
&Graph 10 (RN)                                            & \multirow{2}{*}{0.8125} & {\bf 0.4794}    & 0.4483 & 0.3806                                                      & 6.49                                                & 20.62 \\
& Graph 10 (FF) &    & {\bf 0.5214}    & 0.4848 & 0.4320                                                      & 7.01                                                & 17.14 \\
\cline{2-8} 
&Graph 11 (RN)                                            & \multirow{2}{*}{0.7548} & {\bf 0.3645}    & 0.3472 & 0.2727                                                      & 4.75                                                & 25.19 \\
& Graph 11 (FF) &    & {\bf 0.3989}    & 0.3675 & 0.3104                                                      & 7.88                                                & 22.18 \\
\cline{2-8} 
&Graph 12 (RN)                                            & \multirow{2}{*}{0.6972} & {\bf 0.3231}    & 0.2867 & 0.2560                                                      & 11.26                                                & 20.76 \\
& Graph 12 (FF) &    & {\bf 0.3753 }   & 0.3128 & 0.2195                                                      &16.65                                                & 41.51 \\
\cline{2-8} 
&Graph 13 (RN)                                            & \multirow{2}{*}{0.4213} & {\bf 0.2103}    & 0.2001 & 0.1864                                                      & 4.86 & 11.36 \\
& Graph 13 (FF) &    & {\bf 0.3001}    & 0.2871 & 0.1914                                                      & 4.33                                                & 36.23 \\
\cline{2-8} 
&Graph 14 (RN)                                            & \multirow{2}{*}{0.3125} & {\bf 0.1831}    & 0.1732 & 0.1517                                                      & 5.42  & 17.14 \\
& Graph 14 (FF) &    & {\bf 0.2142}    & 0.1855 & 0.1614                                                      
& 13.40 & 24.67 \\
\cline{2-8} 
&Graph 15 (RN)                                            & \multirow{2}{*}{0.2641} & {\bf 0.1649}    & 0.1373 & 0.1338                                                      & 16.74 & 18.86 \\
& Graph 15 (FF) &    & {\bf 0.2139}    & 0.1941 & 0.1234                                                      
& 9.26  & 42.29 \\
\cline{2-8} 
&Graph 16 (RN)                                            & \multirow{2}{*}{0.6815} & {\bf 0.3994}    & 0.3416 & 0.3246                                                      & 14.48 & 18.74 \\
& Graph 16 (FF) &    & {\bf 0.4664}    & 0.4021 & 0.3632                                                      
& 13.79 & 22.13 \\
\cline{2-8} 
&Graph 17 (RN)                                            & \multirow{2}{*}{0.5122} & {\bf 0.3445}    & 0.2972 & 0.2727                                                      & 13.73  & 20.85 \\
& Graph 17 (FF) &    & {\bf 0.3235}    & 0.3024 & 0.2617
& 6.51   & 19.11 \\
\cline{2-8} 
&Graph 18 (RN)                                            & \multirow{2}{*}{0.4241} & {\bf 0.2971}    & 0.2347 & 0.2160                                                      & 21.01 & 27.29 \\
& Graph 18 (FF) &    & {\bf 0.3209}    & 0.2944 & 0.1747                                                      
& 8.27 & {45.57} \\
\cline{1-8} 
\multicolumn{1}{|c|}{\multirow{12}{*}{\rotatebox{90}{\textbf{Real-world datasets}}}} & Amazon (RN)                                           & \multirow{2}{*}{0.3481}  & {\bf 0.1448}                                                              & 0.1327 & 0.1219 & 9.14 & 18.78                                              \\
\multicolumn{1}{|c|}{}                              & Amazon (FF)     &                                       
& {\bf 0.1962} & 0.1837& 0.1727& 6.79 & 13.59                                            \\
\cline{2-8} 
\multicolumn{1}{|c|}{}                              & DBLP (RN)        & \multirow{2}{*}{0.3249}                                    & {\bf 0.1667}                                                               & 0.1436 & 0.1399                                                        & 16.11                                            & 19.15                                            \\
\multicolumn{1}{|c|}{}                              & DBLP (FF)  &                                          
& {\bf 0.2866}                                                               & 0.2605 & 0.2583 & 10.02                                             & 10.97              \\
\cline{2-8} 
\multicolumn{1}{|c|}{}                              & Youtube (RN)   & \multirow{2}{*}{0.3042}                                          & {\bf 0.1263}                                                               & 0.1179 & 0.1058 & 7.12                                             & 19.34                                          \\
\multicolumn{1}{|c|}{}                              & Youtube (FF)    &                                        
& {\bf 0.1556 }                                                              & 0.1514 & 0.1431 & 2.80                                             & 8.73                                           \\
\cline{2-8} 
\multicolumn{1}{|c|}{} & Facebook (RN)                                           
& \multirow{2}{*}{0.1425}  & {\bf 0.0741}     & 0.0712 & 0.0545 & 3.91 & 26.45                                              \\
\multicolumn{1}{|c|}{} & Facebook (FF)    
&  & {\bf 0.0912} & 0.0878& 0.0745& 3.71 & 18.31                                           
\\
\cline{2-8} 
\multicolumn{1}{|c|}{} & Twitter (RN)                                           
& \multirow{2}{*}{0.1237}  & {\bf 0.0725}     & 0.0701 & 0.0521 & 3.34 & 28.16                                              \\
\multicolumn{1}{|c|}{} & Twitter (FF)    
&  & {\bf 0.0815} & 0.0685& 0.0584& 15.97 & 28.36                                           
\\
\cline{2-8} 
\multicolumn{1}{|c|}{} & Orkut (RN)                                           
& \multirow{2}{*}{0.1249}  & {\bf 0.0824}     & 0.0684 & 0.0301 & 16.99 & {\bf 63.47}                                              \\
\multicolumn{1}{|c|}{} & Orkut (FF)    
&  & {\bf 0.1075} & 0.0845 & 0.0488 & 21.40 & 54.61                                           
\\
\hline
\end{tabular}
}
\egroup
\end{table} 

\subsection{Empirical Evaluation of Complexity}

\begin{figure}[t]
\centering
\includegraphics[height=4.8cm]{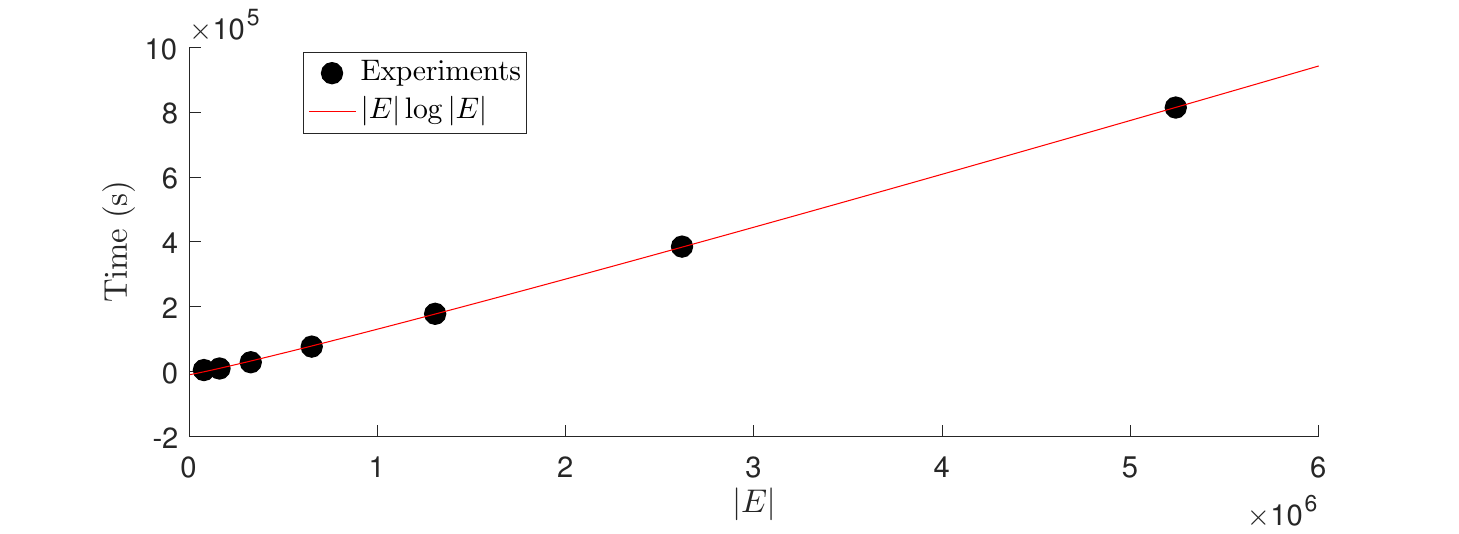}
\caption{The computational complexity of the \textsf{KroMFac} framework.}
\label{fig:complexity}
\end{figure}

We empirically show the average runtime complexity via experiments using synthetic graphs sampled by FF for different numbers of sampled nodes with $N+M = 2^k$ and $k \in \{13,\cdots,19\}$. Then, 30\% of nodes  and their associated edges are deleted by FF sampling to create partially observable networks. Parameters of the LFR benchmark essentially follow those of Graph 1, where MaxC and ON are set proportionally to the number of nodes (refer to Table~\ref{table:1}).\footnote{The runtime complexity under the LFR parameters of other synthetic graphs shows similar trends even if it is not shown in this paper. We note that real-world datasets are not usable for the evaluation of complexity since it is hardly feasible to scale up/down real-world graphs while preserving their structural properties as in the case where synthetic graphs are adopted.} Parameters of the \textsf{KroMFac} framework follow the same settings as in Section~\ref{syndata}. In Figure~\ref{fig:complexity}, we illustrate the plot of the runtime complexity in seconds versus $|E|$, where each point is the average of experimental results obtained by executing the \textsf{KroMFac} process 10 times. An asymptotic curve $|E|\log|E|$ is also shown in the figure, showing a trend that is consistent with our experimental results.
Moreover, we note that since the computation based on a large-size matrix is expensive in terms of memory consumption, it is necessary to adopt cost-effective techniques  (e.g., {\em coordinate format}) to store and compute sparse matrices.

\section{Concluding Remarks}\label{sec:6}
 In this paper, we introduced the problem of discovering overlapping community structures in the context of partially observable networks with both missing nodes and edges. To solve this problem, we developed a novel framework, termed \textsf{KroMFac}, that seamlessly incorporates network completion into community recovery. Specifically, we performed community detection  via regularized NMF based on the Kronecker graph model. In particular, motivated by the insight that adding a proper number of missing nodes and edges to the existing graph would be of significant importance in improving community detection accuracy, we presented how to characterize and select influential nodes via centrality ranking. By adopting the NMI as a performance metric, we validated our proposed \textsf{KroMFac} framework through experiments on both synthetic and real-world datasets. Based on parameter search, we conducted the ablation study for each component in \textsf{KroMFac} and showed that our approach outperforms two baselines by a large margin on synthetic and real-world networks. Additionally, we analytically examined the computational complexity of our framework.

Potential avenues of future research in this area are the inclusion of deep generative graph models for community detection to reduce the inference error even further.

\begin{acks}
This research was supported by the National Research Foundation of Korea (NRF) grant funded by the Korea government (MSIT) (No. 2021R1A2C3004345), by a grant of the Korea Health Technology R\&D Project through the Korea Health Industry Development Institute (KHIDI), funded by the Ministry of Health \& Welfare, Republic of Korea (HI20C0127), and by the Yonsei University Research Fund of 2020 (2020-22-0101). Won-Yong Shin is the corresponding author.
\end{acks}

\section*{Appendix: Effect of Different Centrality Measures}\label{sec:appx}

In this appendix, we empirically compare the effect on the NMI performance along with two centrality measures, namely degree centrality and Katz centrality \cite{katz}, for
analyzing node importance in our approach. While the degree centrality measures the number of immediate neighbors of a node, the Katz centrality measures the influence of a node on its higher-order neighbors at larger distances. It can be seen as a generalization of the eigenvector centrality and penalizes higher-order connections with a factor $\alpha \leq \lambda_{\max}\left({\bf A}_R^{(M)}\right)^{-1}$, where $\lambda_{\max}\left({\bf A}_R^{(M)}\right)$ is the largest eigenvector of ${\bf A}_R^{(M)}$. The Katz centrality of node $u$, denoted by $\text{Cen}_K(u)$, is defined as

\begin{equation}\label{eq:katzcen}
\nonumber
\text{Cen}_{K}(u)=\alpha\sum _{v=1}^{N+M}a_{uv}\text{Cen}_{K}(v)
\end{equation} 
with an initial condition $\text{Cen}_K(u)={c}_\text{init}$ for all $u\in (V \cup V_M)$ and a constant ${c}_\text{init}>0$.
In our empirical experiment, we set the parameter $\alpha$ to $\lambda_{\max}\left({\bf A}^{(M)}_R\right)^{-1}$, which has been shown in~\cite{katzparam} to ensure convergence.

\begin{figure}[t]
\begin{subfigure}[b]{.48\textwidth}
\centering
\includegraphics[width=6.5cm]{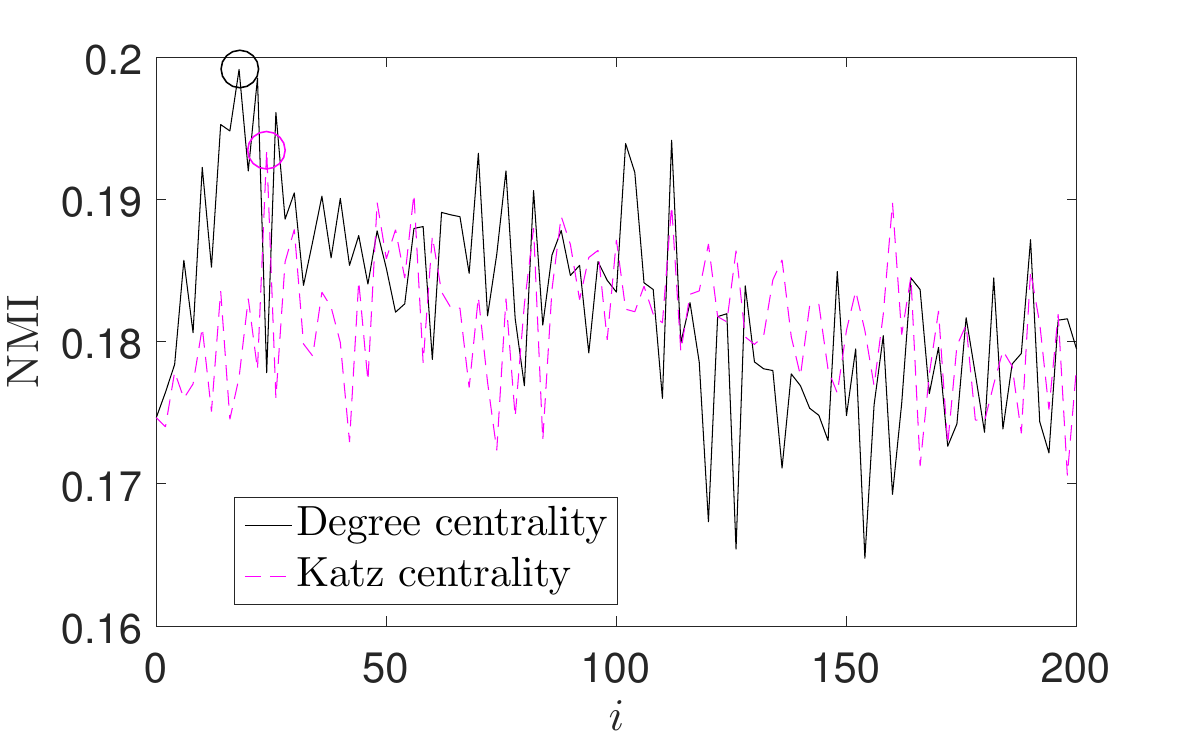}
\caption{RN sampling}\label{fig:comRN}
\end{subfigure}\hspace{1mm}
\begin{subfigure}[b]{.48\textwidth}
\centering
\includegraphics[width=6.5cm]{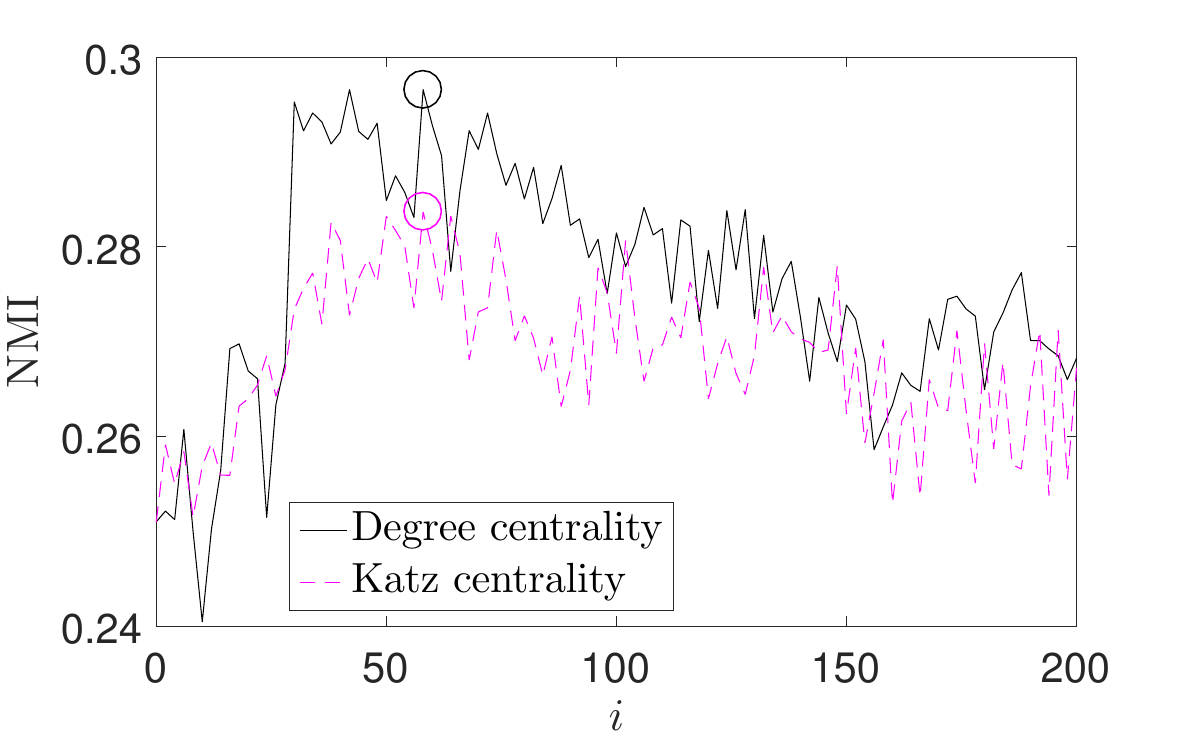}
\caption{FF sampling}\label{fig:comFF}
\end{subfigure}
    \caption{NMI over the number of influential nodes $i$ for
the degree centrality and Katz centrality measures, illustrated on the
example of Graph 2. }
\label{fig:comcen}
\end{figure}

For ease of illustration, we show only the results obtained for Graph 2 in Figure \ref{fig:comcen}, since the other synthetic graphs exhibit the same trend. We plot the NMI against the number $i$ of recovered nodes, when communities are detected for $i$ given nodes that are ranked based on both the degree centrality and the Katz centrality. For RN sampling, the maximum obtained NMI is 0.1991 and 0.1935 at $i$ = 18 and $i$ = 24 when the degree and Katz centrality measures are employed, respectively, as depicted by the empty circles in Figure \ref{fig:comRN}. Similarly, for FF sampling, the maximum NMI is 0.2966 and 0.2832 at $i$ = 58 when the degree and Katz centrality measures are employed, respectively. Clearly, recovered nodes ranked by the degree centrality provide better NMI performance for both RN and FF sampling strategies.

\bibliographystyle{ACM-Reference-Format}
\bibliography{TKDDBib} 

\end{document}